\documentclass[
reprint, 
superscriptaddress,
 amsmath,amssymb,
 aps, prd,
floatfix,
onecolumn,
]{revtex4-2}

\newcommand{\showchanges}{0} 

\usepackage{xcolor}
\if\showchanges1
  \newcommand{\rev}[1]{{\color{red}#1}} 
\else
  \newcommand{\rev}[1]{#1} 
\fi

\usepackage{float}
\usepackage[english]{babel}
\usepackage{subcaption}
\usepackage[letterpaper,top=2cm,bottom=2cm,left=3cm,right=3cm,marginparwidth=1.75cm]{geometry}

\usepackage{amsmath}
\usepackage{graphicx}
\usepackage[colorlinks,citecolor=blue,urlcolor=blue,linkcolor=blue]{hyperref}
\usepackage[dvipsnames]{xcolor}
\usepackage{bm}
\usepackage{amsfonts}
\usepackage[section]{placeins}

\usepackage{array}
\usepackage{scalerel}
\usepackage{tikz}
\usetikzlibrary{svg.path}

\definecolor{orcidlogocol}{HTML}{A6CE39}
\tikzset{
  orcidlogo/.pic={
    \fill[orcidlogocol] svg{M256,128c0,70.7-57.3,128-128,128C57.3,256,0,198.7,0,128C0,57.3,57.3,0,128,0C198.7,0,256,57.3,256,128z};
    \fill[white] svg{M86.3,186.2H70.9V79.1h15.4v48.4V186.2z}
                 svg{M108.9,79.1h41.6c39.6,0,57,28.3,57,53.6c0,27.5-21.5,53.6-56.8,53.6h-41.8V79.1z M124.3,172.4h24.5c34.9,0,42.9-26.5,42.9-39.7c0-21.5-13.7-39.7-43.7-39.7h-23.7V172.4z}
                 svg{M88.7,56.8c0,5.5-4.5,10.1-10.1,10.1c-5.6,0-10.1-4.6-10.1-10.1c0-5.6,4.5-10.1,10.1-10.1C84.2,46.7,88.7,51.3,88.7,56.8z};
  }
}

\newcommand\orcidicon[1]{\href{https://orcid.org/#1}{\mbox{\scalerel*{
\begin{tikzpicture}[yscale=-1,transform shape]
\pic{orcidlogo};
\end{tikzpicture}
}{0}}}}

\begin{document}


\title{Log Gaussian Cox Process Background Modeling in High Energy Physics}

\author{Yuval Frid\,\orcidicon{0000-0003-1565-1773}}
\email[Corresponding author: ]{yuvalfrid@mail.tau.ac.il}
\affiliation{Tel Aviv University, Tel Aviv-Yafo, Israel}

\author{Liron Barak\,\orcidicon{0000-0002-3436-2726}}
\affiliation{Tel Aviv University, Tel Aviv-Yafo, Israel}

\author{Pavani Jairam\,\orcidicon{0000-0003-3600-6925}}
\affiliation{Northwestern University, Evanston, Illinois, USA}

\author{Michael Kagan\,\orcidicon{0000-0002-3386-6869}}
\affiliation{SLAC National Accelerator Laboratory, Menlo Park, California, USA}

\author{Rachel Jordan Hyneman\,\orcidicon{0000-0002-9093-7141}}
\email[Corresponding author: ]{rhyneman@arizona.edu}
\thanks{Currently at University of Arizona. Work began while at SLAC.}
\affiliation{SLAC National Accelerator Laboratory, Menlo Park, California, USA}
\affiliation{University of Arizona, Tucson, Arizona, USA}

\begin{abstract}
Background modeling is one of the most critical components in high energy physics data analyses, and for smooth backgrounds it is often performed by fitting using an analytic functional form. In this paper a novel method based on Log Gaussian Cox Processes (LGCP) is introduced to model smooth backgrounds while making minimal assumptions on the underlying shape. In LGCP, samples are assumed to be drawn from a non-homogeneous Poisson process, with an intensity function drawn from a Gaussian process. Markov Chain Monte Carlo is used for optimizing the hyper parameters and drawing the final fit for the background estimate from the posterior. Synthetic experiments comparing background modeling from functional forms and the LGCP are used to compare the different methods. 
\end{abstract}

\maketitle

\section{Introduction} \label{sec:intro}

In high energy physics, background modeling is the process of estimating the distributions of features for ``background" processes that can mimic a potential signal of interest. 
Background modeling is needed because the data are a mixture of various processes, e.g. they are sampled from a mixture model containing both background processes and potential signal processes of interest. 
By examining populations of event samples, 
one may perform inference on the parameters of interest, i.e. the parameters of a new signal, but a precise understanding of the expected background contributions in a population is critical to performing accurate inference.

More concretely, the collisions of the Large Hadron Collider (LHC)~\cite{Evans:2008zzb} are stochastic processes that produce known Standard Model (SM) particles and possibly also new Beyond the Standard Model (BSM) particles. 
The existence of a BSM signal of interest may manifest in an excess of the event population with a specific signature compared with the expectation from only background SM processes.
As such, one specific strategy to search for BSM particles in LHC data is to exploit the relativistic invariant mass of the decay particles, which
may be tightly constrained around the BSM particle mass. 
The invariant mass of the SM particles produced from SM processes, on the other hand, often follows a smoothly falling spectrum. 
The BSM particle may then be observed as a localized bump on top of a continuous smooth background. 
This strategy exploits the smooth background shape by fitting an analytic functional form to the invariant mass regions on either side of the resonance (denoted the sidebands), then interpolating underneath the resonance to extract the background yield. 
The choice of analytic function is therefore critical to accurately determining the background yield. 
This strategy has been used successfully by both ATLAS~\cite{atlas_tdr} and CMS~\cite{cms_tdr} in a variety of searches for new particles, including in the discovery of the Higgs boson~\cite{ATLAS_Higgs,CMS_Higgs}; in the two-photon decay channel of the Higgs boson, a bump can be observed in the di-photon invariant mass spectrum at the Higgs mass of 125~GeV on top of the roughly exponential smooth di-photon background. 
However, challenges arise in both selecting a function to model the data with sufficient accuracy and in quantifying uncertainties of this procedure, especially when the data set size is small.

In this paper, we present a novel methodology for modeling smooth backgrounds in high energy physics data analyses: the Log Gaussian Cox Process (LGCP)~\cite{LGCP}. 
This method assumes the underlying probability distribution is a non-homogeneous Poisson Process~\cite{poissonProc}, with the intensity function itself being the log transform of a Gaussian Process~\cite{GP}.

\subsection{Motivation}

\begin{figure}[th]
\centering
\includegraphics[width=0.55\linewidth]{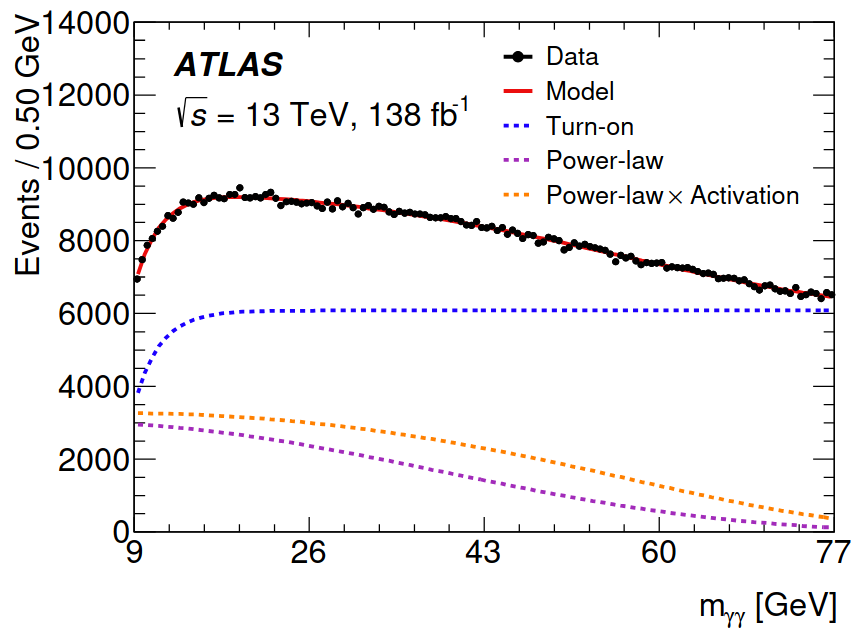}
\caption{\label{fig:Analysis_Fit} Example of a complex background model taken from the very low mass analysis \cite{diphoton}}
\end{figure}

The functional form strategy for fitting smooth backgrounds in physics data analysis carries a number of challenges. 
To better understand these challenges, one can examine how uncertainties related to the choice of functional form are derived in the ATLAS and CMS experiments. 
The ATLAS Experiment assigns uncertainty on the functional form fit based on the so-called spurious signal test, as detailed in Ref.~\cite{ATL-PHYS-PUB-2020-028}. 
A background-only template (representative histogram) is constructed, usually using simulated collision events, and then fit with a signal-plus-background model. 
Any fitted signal arises from the mismodeling by the analytic background function, and it is assigned as an uncertainty on the signal yield. 
The CMS Experiment relies on the discrete profiling method, detailed in Ref.~\cite{Dauncey_2015}, in which an envelope is taken over the negative-log-likelihood curves obtained when using a finite set of test analytic functions. The spurious signal test suffers from a strong dependence on the data set size of the background-only template, since statistical fluctuations in the template may be fitted as spurious signal. 
As a general guidance, the background-only template requires roughly a factor of ten or greater more events than the expected number of background events in data to keep the spurious signal uncertainty reasonable, as detailed in Ref.~\cite{ATL-PHYS-PUB-2020-028}.  
Smoothing methodologies, such as that described in Ref.~\cite{diphoton} utilizing Gaussian Process Regression, may be used to mitigate this statistical inflation. 
This method also assumes that the simulated background template is a realistic description of the true background shape. 
On the other hand, the discrete profiling method assumes a sufficiently large and varied number of analytic functions have been tested to define a realistic envelope on the likelihood. 
In addition to the challenge of deriving suitable uncertainties, the usage of analytic functional forms in background modeling becomes challenging when the background statistics become large enough that higher-order features in the data become statistically significant. 
More degrees of freedom may be required to sufficiently fit the background, but they may also erroneously fit real signal as background. 
Regions where the background is smooth, but not necessarily easily described by few-parameter analytic functions, also present a challenge. For example, the complex background shape on Figure~\ref{fig:Analysis_Fit} (observed in the analysis described in Ref.~\cite{diphoton}) necessitated a functional form with seven floating parameters.

These challenges motivate the exploration of an alternative approach to modeling smooth backgrounds in LHC data. 
One approach is to utilize Gaussian Process Regression (GPR) to fit to the binned sideband and interpolated into the signal region. 
GPR is a Bayesian, non-parametric fitting approach, avoiding the challenge of determining the optimal number of degrees of freedom to both properly describe the background shape while not absorbing signal. 
GPR also does not require an assumption of a specific analytic functional form. 
Assumptions on the shape of the background may be introduced through the choice of the prior mean and the kernel.  
As a Bayesian method, GPR also provides an uncertainty on the fitted posterior mean. 
\rev{This uncertainty is found to generally track with the statistical uncertainty of the sideband data given a motivated choice of kernel and hyperparameters. }

GPR was first proposed for use in background modeling  for high energy physics data analysis in Ref.~\cite{Frate2017ModelingSB}, and has been examined more broadly for applications in collider experiment data analysis in Ref.s~\cite{barr2025gaussianprocessregressionsustainable,Stakia2021AdvancedMA,Mathad_2021,Bertone2016AcceleratingTB,Bertone_2018,Dalmasso2020ConfidenceSA,Kasieczka_2021}. Related methodologies include unfolding~\cite{Bozson2018UnfoldingWG} and extracting signals of BSM physics utilizing the log marginal likelihood~\cite{Yallup2022HuntingFB}.
The ATLAS Experiment has utilized GPR to search for Higgs bosons produced with charm quarks in the diphoton decay channel~\cite{charm_higgs}, and the CMS Experiment has utilized GPR to model the QCD background shape in a search for multijet resonances in scouting data~\cite{Hayrapetyan_2024}, following the statistical procedure outlined in Ref.~\cite{Gandrakota_2023}. ATLAS also explored determining exclusion limit contours with GPs~\cite{ATL-PHYS-PUB-2022-045}.
Beyond the LHC, GPR has been utilized in physics applications for: inferring neutrino oscillation parameters~\cite{Li2020EfficientNO}; smoothing statistical fluctuations in histograms in searches for BSM massless particles~\cite{Ablikim2023SearchFA} and in neutrino cross-section measurements~\cite{Abratenko2024FirstSM,Abratenko2023MeasurementOT,MicroBooNE2024MeasurementOT}; searching for dark matter~\cite{Rodd2024CTAAS}, axions~\cite{Foster2022ExtraterrestrialAS}, and gravitational waves~\cite{Pappas2025HighFrequencyGW}; and machine tuning at a storage ring light source~\cite{Hanuka2019OnlineTA}.

Notably, some limitations on the GPR method may prevent its adoption into wider use in the LHC experiments. 
First, it can only be used for fitting \rev{(input, target)-formatted data, meaning that the count data collected by the LHC experiments must be binned before GPR may be applied. (In contrast, an analytic function may be fit directly to unbinned count data.)} 
Second, GPR assumes that the bin uncertainties are Gaussian, and this assumption is only valid when bins contain sufficient statistics (roughly 10 events per bin or greater). 
This assumption also may lead to biases in the fitted posterior mean for backgrounds with low statistics in all or in a subregion of the sidebands.

The Log Gaussian Cox Proces may offer a promising alternative to either functional forms or GPR. 
Like GPR, the LGCP method provides a flexible, non-parametric fitting framework with posterior uncertainties, unlike the analytic function approach. 
Like analytic functions, though, it may be used to directly fit unbinned data, enabling it to extract additional information from the input dataset, unlike the GPR approach. 
The LGCP may also be used in analyses with limited statistics, where GPR may be biased due to the assumption of Gaussian bin uncertainties. 

\section{Log Gaussian Cox Process Model}

The first assumption for a LGCP is that the probability distribution function (PDF) is described by a non-homogeneous Poisson process \cite{LGCP,gpytorch_lgcp}, resulting in the log likelihood calculated by:
\begin{equation}
\log p(x_1,x_2,\ldots,x_n|\lambda)= \sum_{i=1}^n \log\lambda(x_i) - \int_{X_a}^{X_b} \lambda(x')dx'
\label{eq:log_likelihood}
\end{equation}
Where $\lambda$ is the proposed intensity function, and $X=[x_1, x_2, \ldots, x_n] $ are the input samples with $x_i \in \mathbb{R}^{[X_a,X_b]}$. 
For LHC analyses, each input sample represents the value of the variable-of-interest (such as invariant mass) for a recorded event. 
For simplicity, the samples are transformed linearly to a new $x$-range of [0,1]. 
The second assumption for a LGCP is that the intensity function itself is a stochastic process, where $\lambda(x)$ is the log transform of a Gaussian Process: 
\begin{equation}
\lambda(x) = N_E \cdot \exp^{Z(x)}
\label{eq:lambda}
\end{equation}
with $Z(x) \sim GP(\mu(x), K(x,x'))$ and $GP(\mu(x),  K(x,x'))$ denoting a Gaussian process with mean $\mu(x)$ and covariance kernel $K(x,x')$ (i.e. the joint distribution of any finite vector $(Z(x_1), ... , Z(x_n))$ is Gaussian).
This means that $\lambda(x)$ is equal to the total amount of samples $N_E$ multiplied by the exponential of a Gaussian process $Z(x)$. More explicitly, this identity defines the prior of the log intensity function over a finite set of measurements $[x_1, ... , x_n]$ as a multivariate normal distribution:

\begin{equation}
p \left( \begin{bmatrix} \log \frac{\lambda(x_1)}{N_E} \\ \vdots \\ \log \frac{\lambda(x_n)}{N_E} \end{bmatrix} \right) = \mathcal{N} \left(\begin{bmatrix} \mu(x_1) \\ \vdots \\ \mu(x_n) \end{bmatrix},  \begin{bmatrix} K(x_1, x_1) &  ... & K(x_1, x_n) & \\ \vdots & & \vdots \\ K(x_n, x_1) & ... & K(x_n, x_n) \end{bmatrix} \right)
\label{eq:log_prior}
\end{equation}

Using Bayes' formula, along with the log likelihood \rev{Equation}~\ref{eq:log_likelihood} and the log prior \rev{Equation}~\ref{eq:log_prior}, the posterior $p(Z(x) | X,\Theta_{\mu,\Sigma})$ can be calculated, with $X$ as the set of samples, and $\Theta_{\mu,\Sigma}$ the parameters defining the multivariate Z:
\begin{equation}
p(Z(x) | X, \Theta_{\mu,\Sigma}) = \frac{p(X|Z(x)) \cdot p(Z(x) | \Theta_{\mu,\Sigma})}{p(X | \Theta_{\mu,\Sigma})}
\label{eq:posterior}
\end{equation}
The marginal likelihood $p(X | \Theta_{\mu,\Sigma})$ is the probability of sampling the set X given the Log Gaussian parameter set $\Theta_{\mu,\Sigma}$, and is used later for hyper-parameter optimization. It is calculated by integrating over all possible $Z(x)$ sampled from $GP(\mu(x), K(x,x'))$, described in Equation~\ref{eq:marginal_likelihood}.
\begin{equation}
p(X | \Theta_{\mu,\Sigma}) = \int_{Z(x)}{p(X| Z(x))\cdot p(Z(x) | \Theta_{\mu,\Sigma}) \cdot dZ}
\label{eq:marginal_likelihood}
\end{equation}

Due to the complexity of integrating over all possible intensity functions $Z(x)$, a Monte Carlo integration \rev{Equation}~\ref{eq:Monte_Carlo_Integration} is used instead to estimate the marginal likelihood:

\begin{equation}
 \int{F(x)\cdot p(x) \cdot dx} \approx \frac{1}{N_{MC}}\sum_{i= 1}^{N_{MC}} F({x_i})
\label{eq:Monte_Carlo_Integration}
\end{equation}

where $N_{MC}$ different samples are generated from $p(x)$, and their mean $F(x)$ value is the estimation of the integral.
Plugging Equation~\ref{eq:Monte_Carlo_Integration} into Equation~\ref{eq:marginal_likelihood} results in the following:

\begin{equation}
p(X | \Theta_{\mu,\Sigma}) = \int_{Z(x)}{p(X| Z(x))\cdot p(Z(x) | \Theta_{\mu,\Sigma}) \cdot dZ}  \approx \frac{1}{N_{MC}}\sum_{i= 1}^{N_{MC}} p({X|Z_i(x)})
\label{eq:Marginal_Integration}
\end{equation}

For each set of hyper-parameters $\Theta_{\mu,\Sigma}$, a large amount of $Z(x)$ functions are generated, and the mean of their likelihood values is the approximation of the marginal likelihood.
The general GP is simplified by setting the mean $\mu$ to a 0 vector, and the $\Sigma$ covariance matrix to the Radial Basis Function (RBF) kernel (also referred to as the Squared Exponential kernel)  $\mathcal{K}$ defined as 
\begin{equation}
 \rev{\mathcal{K} (x,x'|\ell,\sigma^2) = \sigma^2 \exp^{\frac{-(x-x')^2}{2\ell^2}}}
\label{eq:RBF_Kernel}
\end{equation}
where $\sigma$ and $\ell$ are the two hyperparameters. 
The former is a normalization factor, while the latter -- the length scale -- encodes the minimum feature size in the data. 
The kernel assumptions are therefore physically interpretable.

The intensity function $\lambda$ from \rev{Equation}~\ref{eq:lambda} is now defined entirely by a GP with two positive hyper-parameters:
\begin{equation}
\rev{Z \sim \mathcal{N}(0,\mathcal{K}({x,x'|\ell,\sigma^2}))}
\label{eq:RBF}
\end{equation}

The values for $\ell,\sigma^2$ are optimized by maximizing the Monte-Carlo integral in Equation~\ref{eq:Monte_Carlo_Integration}

\section{Inference with Markov Chain Monte Carlo}

\begin{figure}
\centering
\begin{subfigure}{0.48\linewidth}
    \includegraphics[width=\linewidth]{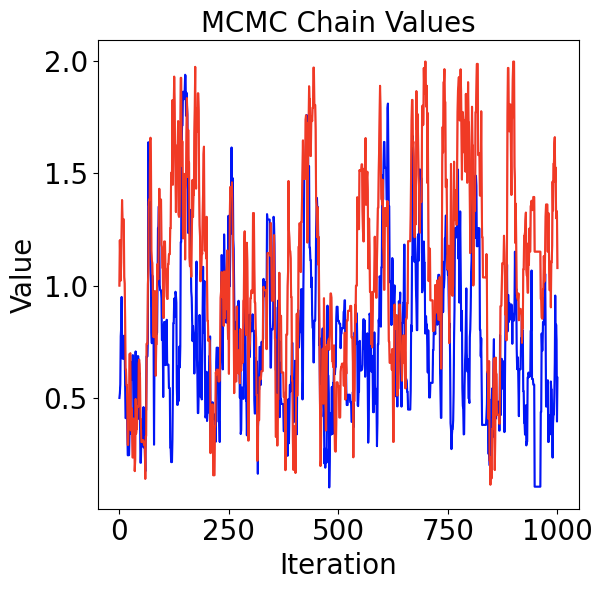}
\end{subfigure}
\hfill
\begin{subfigure}{0.48\linewidth}
    \includegraphics[width=\linewidth]{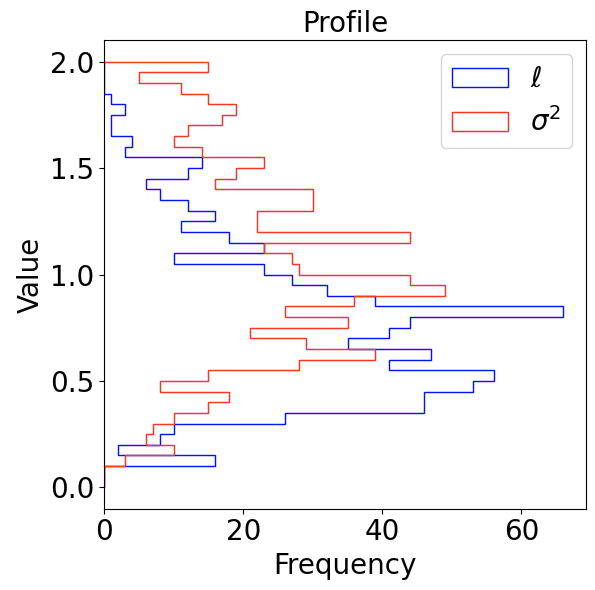}
\end{subfigure}
\caption{\label{fig:HP_MCMC} Examples of the MCMC values (left) and profiles (right) for the $\ell$ (blue) and $\sigma^2$ (red) hyper-parameters. The means of the histograms on the right are the optimized hyper-parameters.}
\end{figure}

Inferring the intensity function $\lambda(x)$ using the input samples is performed in two stages using Markov Chain Monte Carlo (MCMC). In the first stage, the hyper-parameter values for the GP are optimized using the Metropolis-Hastings (MH) \cite{MCMC} algorithm. A 2D vector $\bm{\Theta_0}$ with initial values $\ell_0,\sigma^2_0$ is selected, and the marginal likelihood $p(X|\bm{\Theta_0})$ is calculated by sampling \rev{10,000} different $Z(x)$ functions from the GP and inserting into Equation~\ref{eq:Marginal_Integration}. Each iteration, a small step is drawn from a narrow normal distribution, and a new marginal likelihood is calculated. The acceptance ratio $\mathcal{A}$ for the $N^{th}$ step is defined as

\begin{equation}
\mathcal{A} = \frac{p(X|\bm{\Theta_n})}{p(X|\bm{\Theta_{n-1}})}
\label{eq:Acceptance Ratio}
\end{equation}

and the next value in the chain is determined by:

\begin{equation}
\bm{\Theta_n} = 
\begin{cases}
    \bm{\Theta_n} & \text{if } \mathcal{A} > u, u  \sim\mathcal{U}(0,1)  \\
    \bm{\Theta_{n-1}} & \text{else }  \\
\end{cases}
\label{eq:MCMC_Step}
\end{equation}

where $u$ is a value drawn from the uniform distribution $\mathcal{U}$ between 0 and 1.
These chains estimate the probability distributions for the hyper-parameters $\ell$ and $\sigma^2$.
The mean values for $\ell$,$\sigma^2$ are taken as the optimized hyper-parameters for the RBF. 
Examples of the hyper-parameter MCMC chains are illustrated in Figure~\ref{fig:HP_MCMC}.

\begin{figure}[ht!]
\centering
\includegraphics[width=0.7\linewidth]{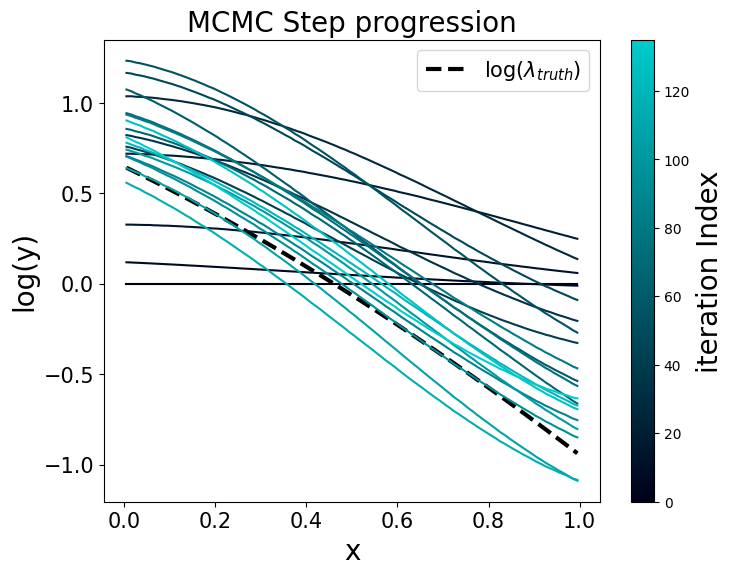}
\caption{\label{fig:Posterior_MCMC} Progression of the posterior MCMC. Each curve represents $Z_i(x)$ for iteration $i$ in the MCMC, defining the posterior distribution $p(Z(x)|X,\ell_o,\sigma^2_o)$}
\end{figure}

Once the RBF kernel is optimized, the next step is to create a new MCMC chain for $Z(x)$. This chain determines the final background fit based on the posterior (Equation~\ref{eq:posterior}.) 
The initial value in the chain is a 0 vector. Each iteration, a step $Z(x)'$ is sampled from $\mathcal{N}(0,\mathcal{K}(\ell,\sigma^2))$, and the next value in the chain is defined as $Z_n(x)$ = $Z_{n-1}(x) + \epsilon \cdot Z(x)'$, where $\epsilon$ is a small scaling, analogous to learning rate in other machine learning algorithms. Similarly to the previous MCMC for the hyper-parameters, the acceptance ratio $\mathcal{A}$ determines the next step:

\begin{equation}
\mathcal{A} = \frac{p(Z_n(x)|X)}{p(Z_{n-1}(x)|X)} = \frac{p(X|Z_n(x)) \cdot p(Z_n(x)|\ell_o,\sigma^2_o)}{p(X|Z_{n-1}(x)) \cdot p(Z_{n-1}(x)|\ell_o,\sigma^2_o)} 
\label{eq:Acceptance Ratio lambda}
\end{equation}

with $\ell_o$ and $\sigma^2_o$ the optimized hyper-parameters, and 

\begin{equation}
Z_n(x) = 
\begin{cases}
    Z_n(x) & \text{if } \mathcal{A} > u, u \in \mathcal{U}(0,1)  \\
    Z_{n-1}(x) & \text{else }  \\
\end{cases}
\label{eq:MCMC_Step_lambda}
\end{equation}
The progression of the $Z(x)$ values is shown in Figure~\ref{fig:Posterior_MCMC}.
As before, this chain of vectors $Z_n(x)$ estimate the probability distribution $p(Z(x)|X,\ell_o,\sigma^2_o)$.
For the final fit, the values of the $p(Z(x)|X,\ell_o,\sigma^2_o)$ chain for each $x_i$ are used separately. The median value of the chain distribution for each $x_i$ is the result for $Z(x_i)$, with the 16 and 84 percentiles taken as the $1\sigma$ uncertainty bands, analogous to a normal distribution. The first 20\% of the MCMC iterations are tossed, and only the last 80\% are used to determine the final result. This method is therefore an approximation of a true LGCP method, where the underlying intensity function is modeled with a Gaussian Process controlled by a set of inducing points. Here the final result is not a Gaussian Process, but a distribution derived by iteratively adding small (x,y) vectors derived from a Gaussian Process. This method was used because of the high correlation observed between inducing points, which restricts the MCMC convergence of the LGCP.

\section{Background + Signal Fit}\label{sec:SS}

In order to quantify the performance of the LGCP method, fits are performed to toy datasets containing a background-only-like distribution of events as well as a signal-plus-background distribution of events. 
For LGCP combined signal-plus-background fits, an additional hyperparameter, $N_S$, representing the number of signal samples, is included in the MCMC optimization.
The intensity function is now defined as:
\begin{equation}
\lambda(x)= (N_E-N_s) \cdot \exp^{Z(x)} + N_s \cdot S(x)
\label{eq:lambda_SS}
\end{equation}
where $S(x)$ is the known signal PDF (a Gaussian distribution for these results). The value for $N_S$ is estimated through the MCMC chain for the RBF hyper-parameters $\ell,\sigma^2$. 

As a comparison against the LGCP, fits are also performed using a GP with an RBF kernel, defined using the \texttt{scikit-learn} package~\cite{scikit-learn}. 
For the GP fit, a localized signal kernel described in Ref.~\cite{Frate2017ModelingSB} is added to the kernel to account for any fitted signal: 
\begin{equation}
    K(x,x')=\sigma_s^2e^{-\frac{1}{2\ell_s^2}(x-x')}e^{-\frac{1}{2t^2}((x-m)^2+(x'-m)^2)}
\end{equation}
where $\ell_s$ denotes the signal lengthscale, $m$ denotes the signal's location in $x$ (the mass), and $t$ denotes the width of the signal. 
For this result, $\ell_s$ and $\sigma_s$ are optimized using the log marginal likelihood, while $m$ and $t$ are fixed to match the expected signal being fit.

\section{Results}

\subsection{Generated Samples}

Toy datasets were generated from different analytic functions and with different statistics to compare the performance of the LGCP model against other fitting methods. 
In this study, the following test functions were used to describe two general cases of a background pdf:
\[F_{1}(x|a,b) = \exp^{-\frac{(1+x)^a}{x+b}} \]
\[F_{2}(x|a,b,c) = \frac{\exp^{-a \cdot x}}{1+\exp^\frac{b-x}{c}} \]
$F_1$ describes a smoothly falling distribution, similar to the diphoton invariant mass spectrum which might be encountered in a measurement of the Higgs boson in the diphoton decay channel. 
Toy datasets were generated using the parameters $a = 2.5$ and $b= 1.5$.
$F_2$ describes a distribution starting with a turn on (such as may be induced by a trigger threshold effect) with a smoothly falling tail. 
This shape illustrates a drop in efficiency when reaching the lower end of the $x$-spectrum, similarly to the analysis referenced in Figure~\ref{fig:Analysis_Fit}. 
Toy datasets were generated using the parameters $a=2$, $b= 0.1$, and $c= 0.05$.



\begin{figure}[h!]
\centering
\begin{subfigure}{0.48\linewidth}
    \includegraphics[width=\linewidth]{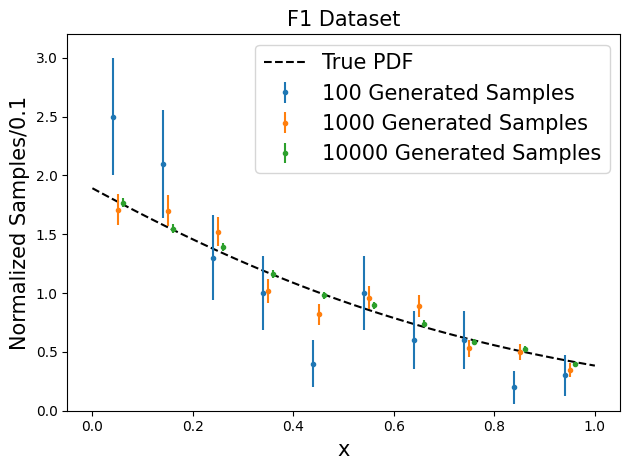}
\end{subfigure}
\hfill
\begin{subfigure}{0.48\linewidth}
    \includegraphics[width=\linewidth]{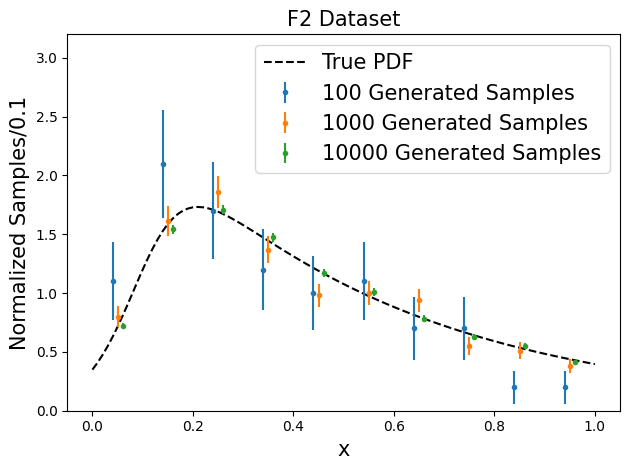}
\end{subfigure}
\caption{\label{fig:Toy_Dataset} Examples of toy datasets generated from $F_1$ and $F_2$ with different statistics ($100$, $1000$, and $10,000$ events). The toy datasets are normalized for shape comparison. The dashed line shows the true value of the analytic function (left: $F_1$, right: $F_2$) used to generate the toy datasets.}
\end{figure}

All tests were performed in the range [0,1] and can be generalized to any range by a simple linear transformation. 
Three working points were determined, with $100$, $1000$ and $10,000$ events per dataset. 
For each working point, 1000 toy datasets were generated using different random seeds. 
These functional forms, along with examples of toy datasets, are shown in Figure~\ref{fig:Toy_Dataset}. 

\subsection{Background Modeling}

Each toy dataset is fit with the LGCP, an unbinned Maximum Likelihood Estimator (MLE), and a GP. 
The LGCP for $F_1$ uses the RBF kernel, while $F_2$ requires a Gibbs kernel, due to the sharp turn on in the lower end. The Gibbs kernel assumes the length scale isn't constant, and in the case of $F_2$ a linear function was chosen, increasing the total hyper-parameters from 2 to 3:
 \begin{equation}
  K(x,x') = \sigma^2 \exp^{\frac{-(x-x')^2}{\ell(x)^2+\ell(x')^2}}, \ell(x) = \ell_0+\ell_1\cdot x
 \label{eq:Gibbs_Kernel}
 \end{equation}

The MLE fit was done twice, once with a functional form similar, but not identical to, the true background shape (denoted as the ``estimated MLE’’), and another with the same functional form used for sample generation (denoted as the ``optimal MLE''). The functional forms for the estimated MLEs are:
\[F_{1}^{est}(x|a,b) = \exp^{-(a \cdot x)^b} \]
\[F_{2}^{est}(x|a,b,c) = \frac{x^b}{c^b+x^b} \cdot \exp^{-a \cdot x} \]
The former illustrate a more realistic result of a physics analysis, where the true underlying background shape is not known exactly while the latter is theoretically the ideal case.  

For the GP fits, the data is binned into 10, 100, or 1000 equal-size bins for the toys containing 100, 1000, or $10,000$ events, respectively, and the bin content is used as the diagonal variance (emulating a Poisson variance). 
Similarly to LGCP, the RBF kernel is used for $F_1$, while $F_2$ requires a Gibbs kernel with a linear function.

\begin{figure}[H]
\centering
\begin{subfigure}{0.32\linewidth}
    \includegraphics[width=\linewidth]{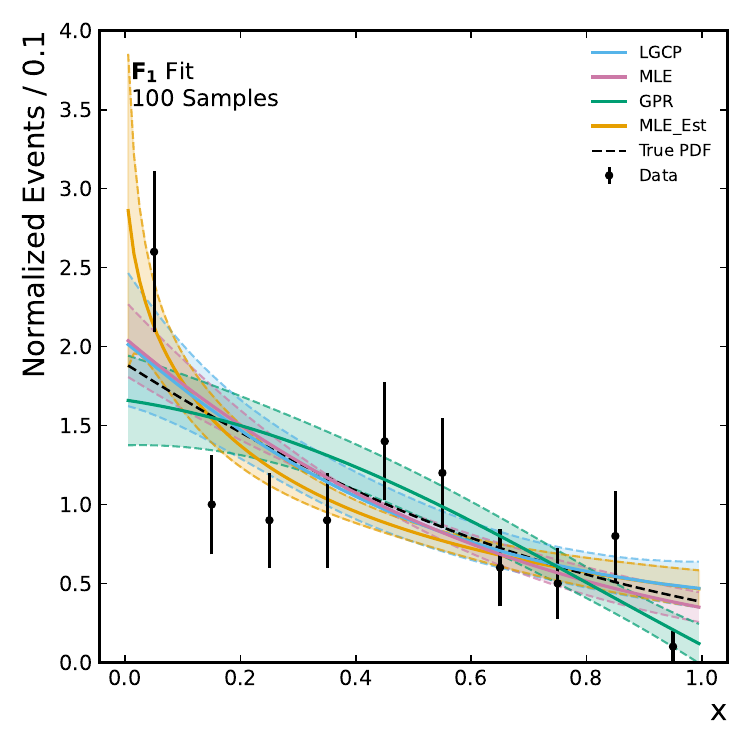}
\end{subfigure}
\hfill
\begin{subfigure}{0.32\linewidth}
    \includegraphics[width=\linewidth]{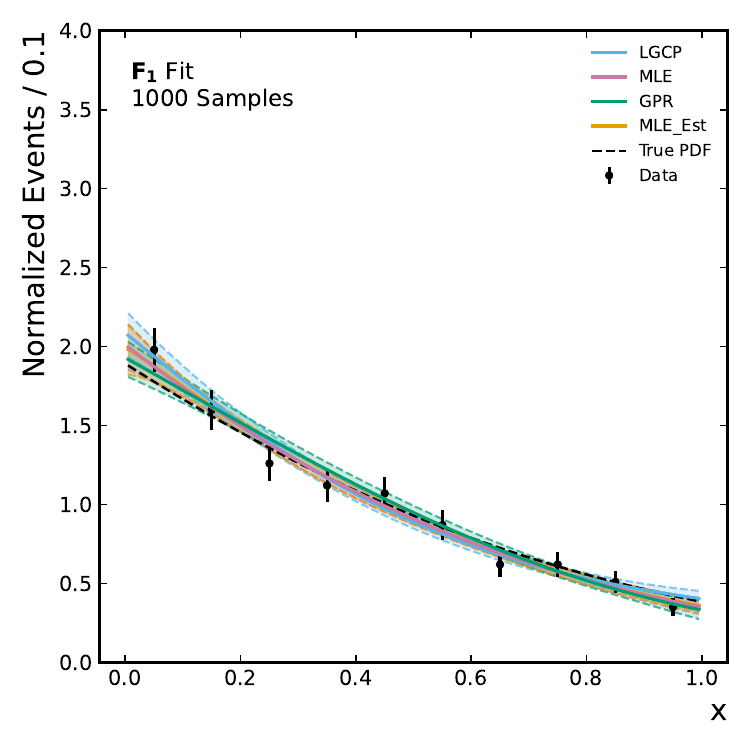}
\end{subfigure}
\hfill
\begin{subfigure}{0.32\linewidth}
    \includegraphics[width=\linewidth]{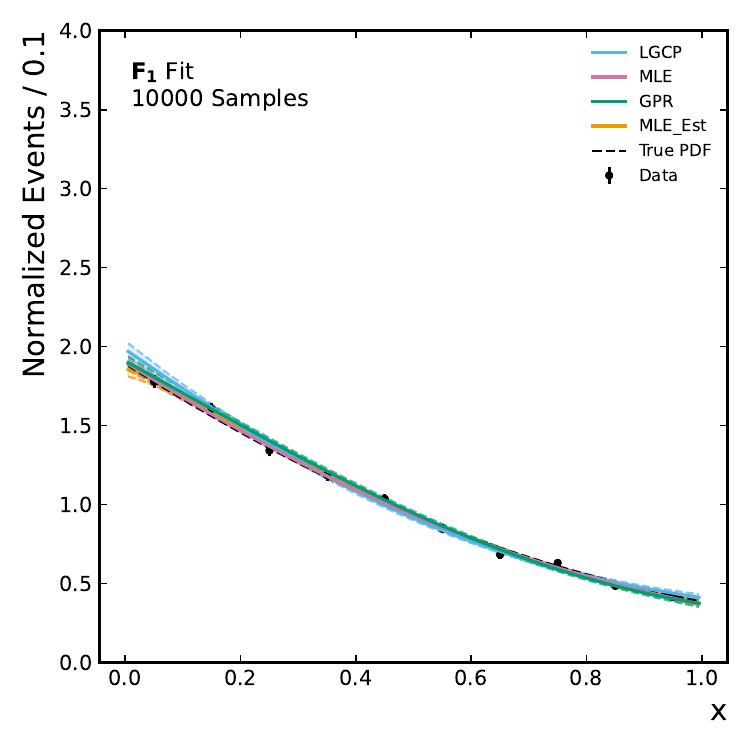}
\end{subfigure}
\caption{\label{fig:F1_Fit_Example} Fits (LGCP in blue, GP in green, estimated MLE in orange, and optimal MLE in pink) to an example toy dataset (shown by the black points) with: (left) 100 events, (middle) 1000 events, and (right) 10,000 events. The shaded band denotes the fit uncertainty from each method. The dashed black line denotes the true shape of the function $F_1$ used to generate the toy dataset.}
\end{figure}

\begin{figure}[H]
\centering
\begin{subfigure}{0.32\linewidth}
    \includegraphics[width=\linewidth]{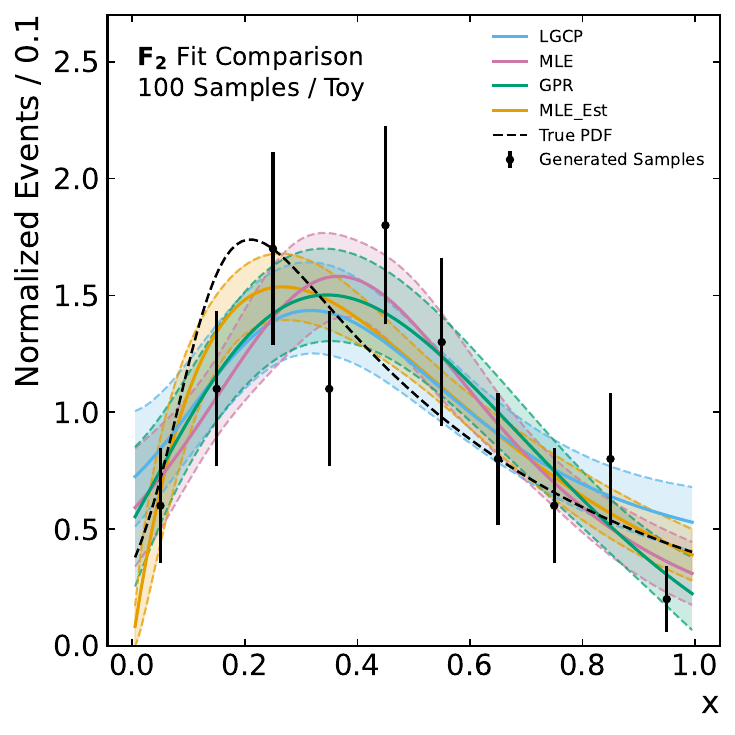}
\end{subfigure}
\hfill
\begin{subfigure}{0.32\linewidth}
    \includegraphics[width=\linewidth]{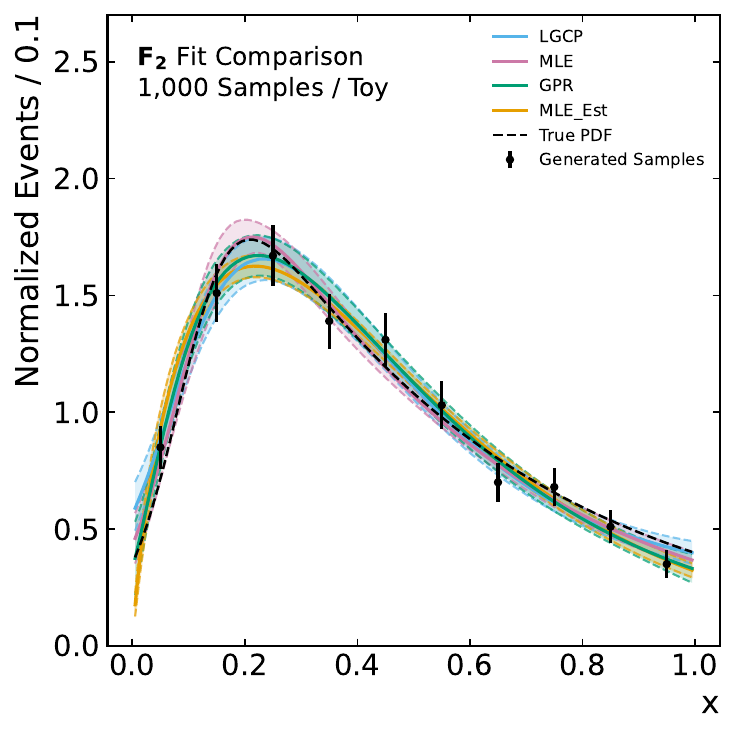}
\end{subfigure}
\hfill
\begin{subfigure}{0.32\linewidth}
    \includegraphics[width=\linewidth]{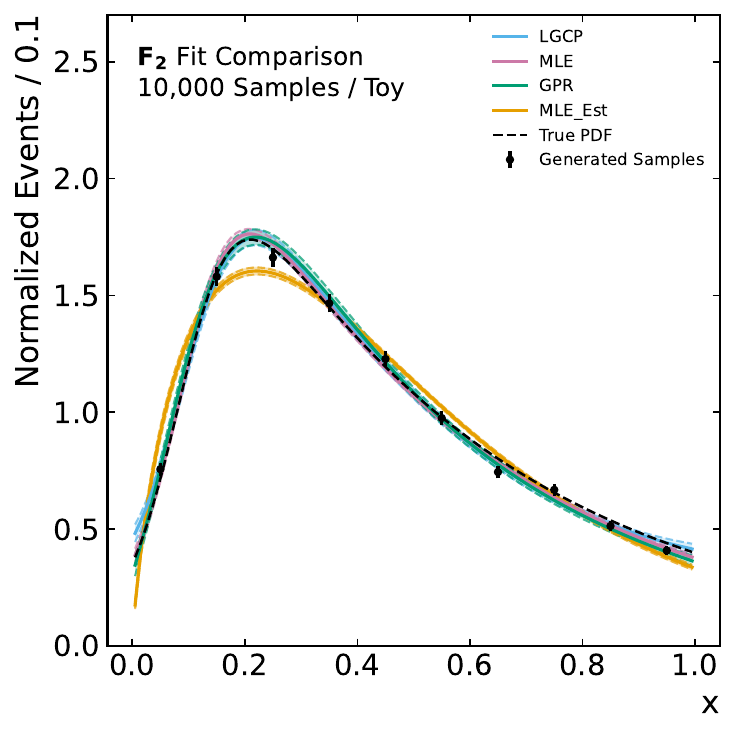}
\end{subfigure}
\caption{\label{fig:F2_Fit_Example} Identical to Figure~\ref{fig:F1_Fit_Example}, with $F_2$ as the underlying PDF.}
\end{figure}

Examples of fits to toy datasets generated from $F_1$ and $F_2$ are shown in Figures~\ref{fig:F1_Fit_Example} and \ref{fig:F2_Fit_Example}, respectively.
As the number of events increases, the optimal MLE improves on average, and uncertainties decrease. For the estimated MLE, the observed behavior is less consistent, and the more complex shape of $F_2$ requires a more flexible fit to compensate for the difference in functional forms.
For the LGCP fit, the mean improves with increasing statistics, but the uncertainties could be considered over-constrained as statistics increase. 
A similar trend is also observed for the GPR method.

For each working point, the 1000 toy dataset fits were used to generate a pull plot. For each $x$ value in each fit, the distribution $Pull(x)$ is determined using the following formula:

\begin{equation}
    Pull(x) = \frac{F_{Fitted}(x)-F_{Truth}(x)}{\delta F_{Fitted}(x)}
\label{eq:Pull}
\end{equation}
where $F_{Fitted}(x)$ denotes the fit to the toy dataset, $F_{Truth}(x)$ denotes the functional form ($F_1$ or $F_2$) used to generate the toy dataset, and $\delta F_{Fitted}(x)$ denotes the $\pm1\sigma$ uncertainty of the fit.
The mean and standard deviation of $Pull(x)$ for each $x$ determine the value and uncertainties for a given $x$.
Ideally, the pulls will demonstrate a mean of $0$ and $\pm1\sigma$ band of 1 for all $x$, meaning both the fit and uncertainty values are well determined. 

\begin{figure}[H]
\centering
\begin{subfigure}{0.32\linewidth}
    \includegraphics[width=\linewidth]{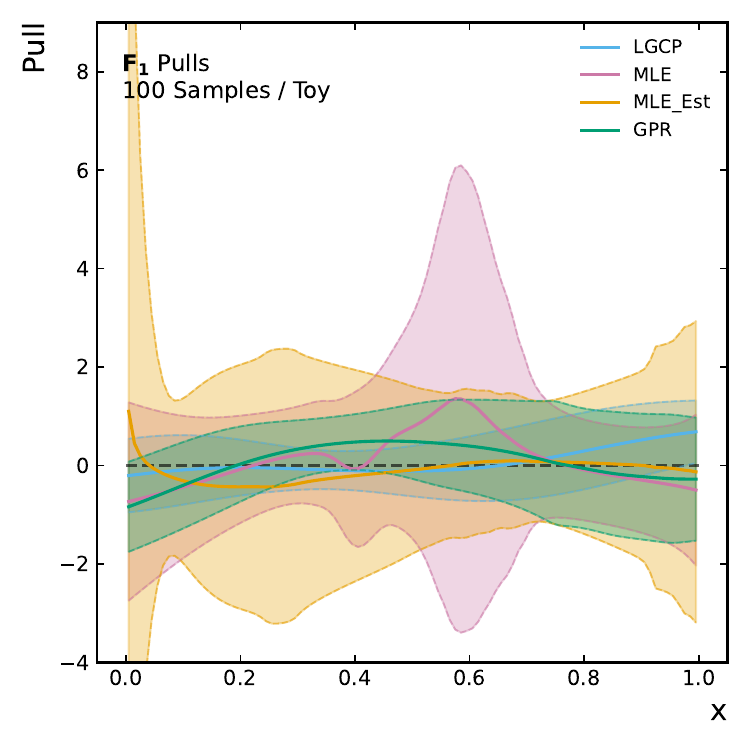}
\end{subfigure}
\hfill
\begin{subfigure}{0.32\linewidth}
    \includegraphics[width=\linewidth]{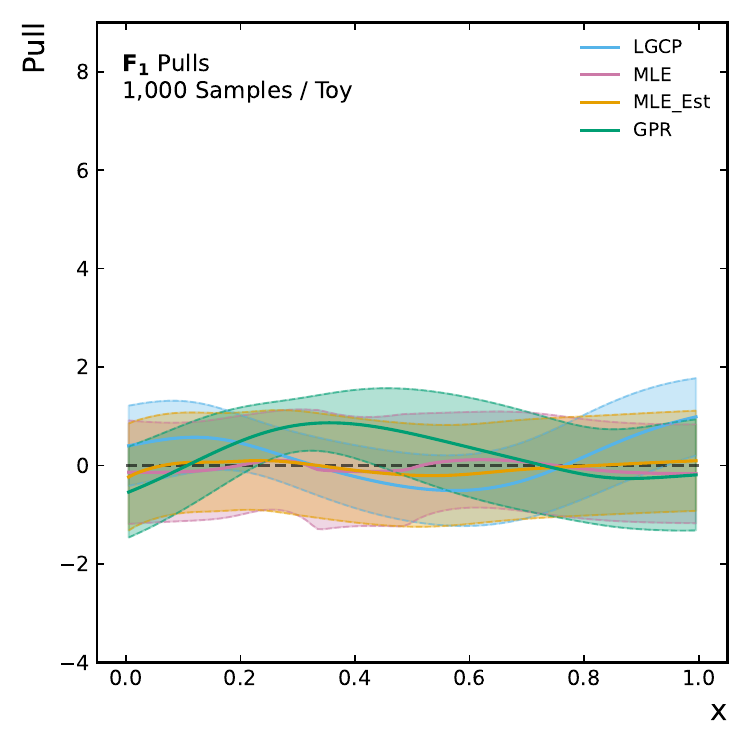}
\end{subfigure}
\hfill
\begin{subfigure}{0.32\linewidth}
    \includegraphics[width=\linewidth]{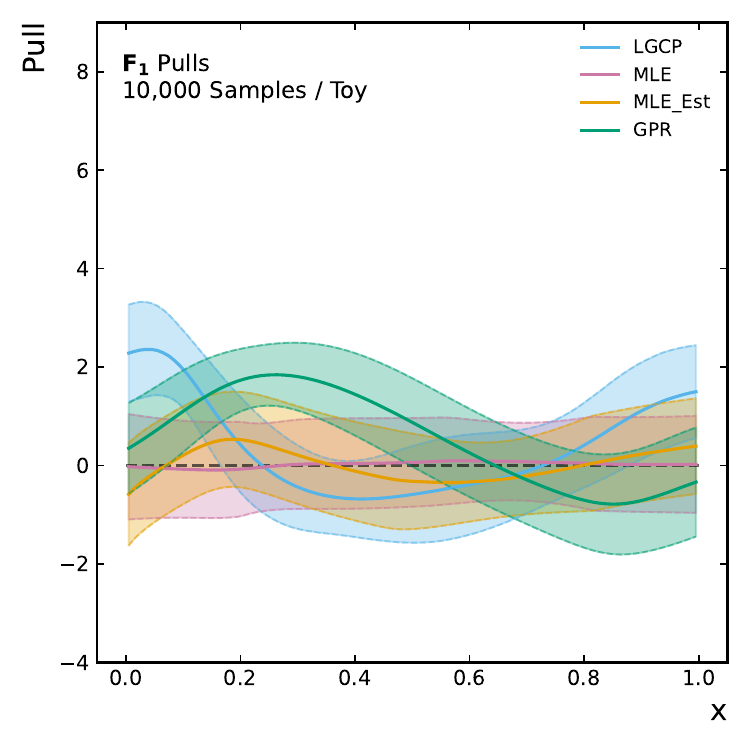}
\end{subfigure}
\caption{\label{fig:F1_Pull_Plots} The average of the pulls from the (blue) LGCP, (green) GP, (orange) MLE Estimate, and (pink) MLE fits to the $1000$ toy datasets generated from $F_1$. The shaded band denotes the $\pm1\sigma$ range of the pulls at each point in $x$.}
\end{figure}

The pulls are shown in Figure~\ref{fig:F1_Pull_Plots} for the LGCP, GPR, optimal MLE, and estimated MLE when fitting $F_1$. 
For the lowest-statistics datasets (100 events per toy dataset), the LGCP and GPR perform similarly well, and both appear to out-perform the MLEs. 
The MLE using the toy-generation function is especially unstable when fitting very-low-statistics toys. For the mid-statistics datasets (1000 events per toy dataset), the LGCP pulls perform comparably to those of the GPR. 
Both show a mild bias larger than that of the MLEs, which both perform very well with minimal bias and near-ideal uncertainties. 
For the highest-statistics datasets tested (10,000 events per toy dataset), the average of the LGCP pulls deviates greater than 1 sigma away from the true value, indicating bias, near the edges of the $x$-range, though near the middle of the $x$-range, its pulls are generally smaller than the GPR. 
For all tested statistical regimes, the uncertainty bands for the LGCP and GPR fits are frequently smaller than $\pm 1$, indicating the uncertainties are under-estimated from these methods. 

\begin{figure}[H]
\centering
\begin{subfigure}{0.32\linewidth}
    \includegraphics[width=\linewidth]{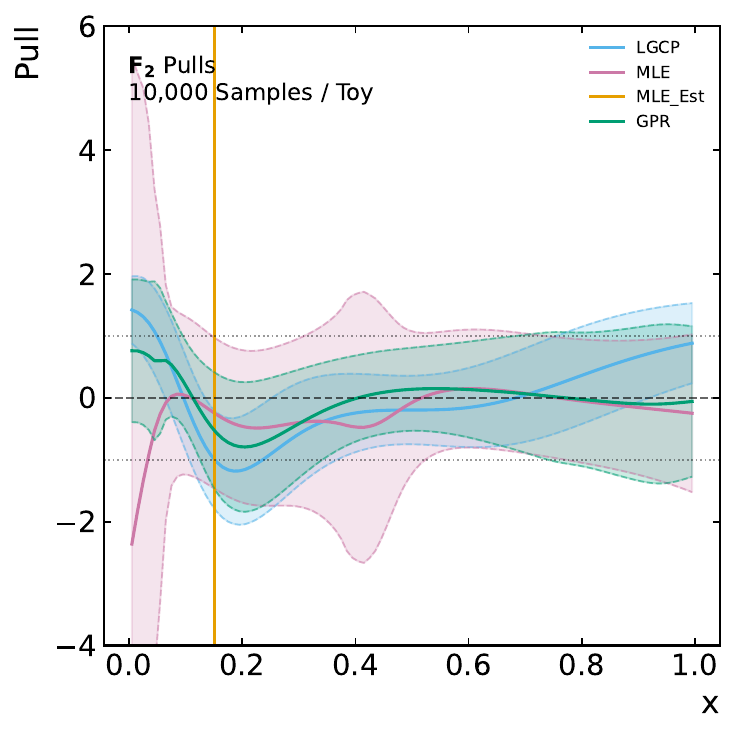}
\end{subfigure}
\hfill
\begin{subfigure}{0.32\linewidth}
    \includegraphics[width=\linewidth]{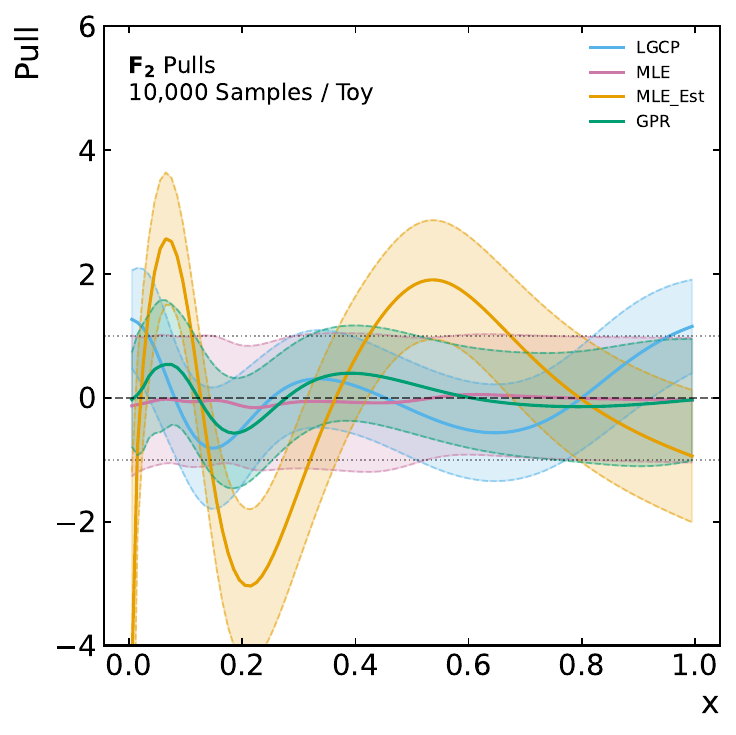}
\end{subfigure}
\hfill
\begin{subfigure}{0.32\linewidth}
    \includegraphics[width=\linewidth]{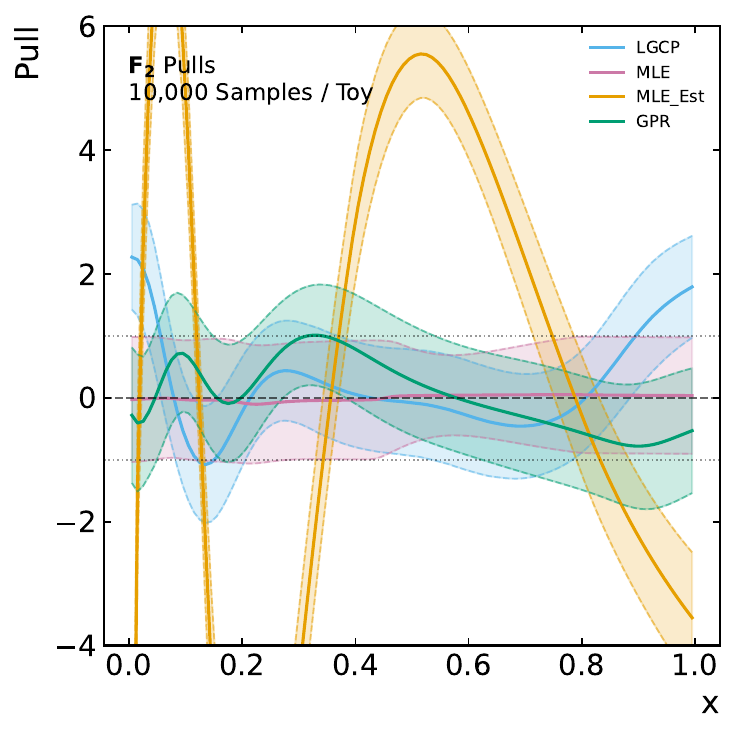}
\end{subfigure}
\caption{\label{fig:F2_Pull_Plots} The average of the pulls from the (blue) LGCP, (green) GP, (orange) MLE Estimate, and (pink) MLE fits to the $1000$ toy datasets generated from $F_2$. The shaded band denotes the $\pm1\sigma$ range of the pulls at each point in $x$.}
\end{figure}

Results when fitting $F_2$ are shown in Figure~\ref{fig:F2_Pull_Plots}. 
For all statistics regimes, the estimated MLE demonstrates large pulls and poorly-estimated uncertainties, despite the overall fit quality appearing adequate in the low- and mid-statistics regimes shown in Figure~\ref{fig:F2_Fit_Example}. 
The LGCP and GPR therefore may significantly outperform the MLE in scenarios in which the choice of analytic function is sub-optimal. 
For the lowest-statistics datasets, the LGCP, GPR, and MLE perform similarly, with bias observed from all three in the low-$x$ region consistent with challenges modeling the turn-on feature in datasets where it is poorly resolved. 
For the mid- and high-statistics datasets, the LGCP and GPR exhibit a comparable mild bias compared with the MLE in the middle of the $x$-range, with the LGCP exhibiting larger biases near the edges of the range. 
The high-statistics dataset pulls from both $F_1$ and $F_2$ suggest that the LGCP suffers from greater edge distortions than the GP or analytic function fits. 
However, this effect may be mitigated in a physics analysis by fitting a larger side-band region than needed, then truncating the biased edges. 

\FloatBarrier
\subsection{Spurious Signal}

A spurious signal test (as described in Section~\ref{sec:intro}) is performed for each toy dataset to further quantify any bias from each method. 
For these results, no signal is injected into the toy datasets, and so they represent a background-only template which may be used in a physics analysis. 
Due to statistical fluctuations in the toy datasets, some non-zero spurious signal is expected in individual toy dataset fits. 
A non-zero mean of the average fitted spurious signal over all toy datasets quantifies the mismodeling between the fit method and the underlying functional form used to generate the toy datasets. 

\begin{figure}[H]
\centering
\begin{subfigure}{0.32\linewidth}
     \includegraphics[width=\linewidth]{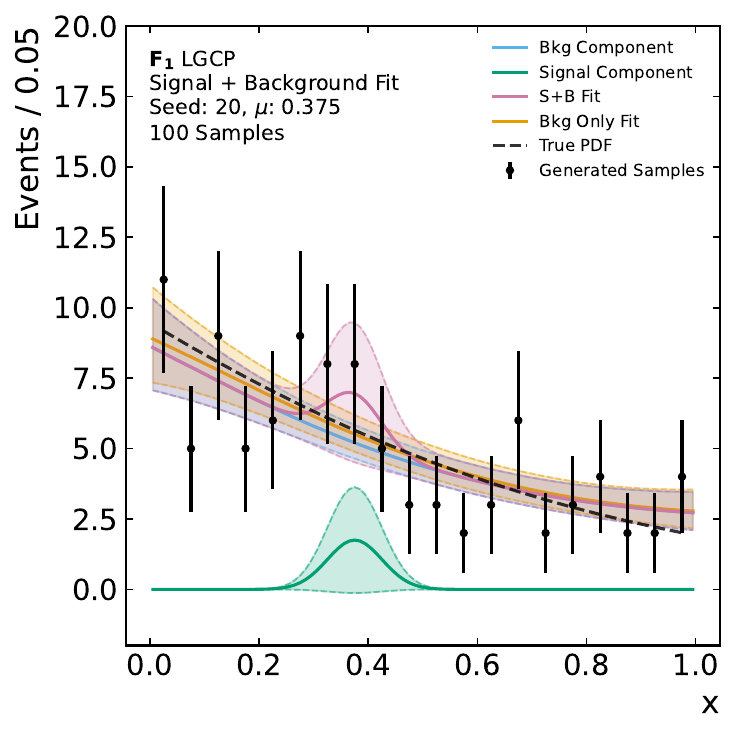}
\end{subfigure}
\begin{subfigure}{0.32\linewidth}
\hfill
     \includegraphics[width=\linewidth]{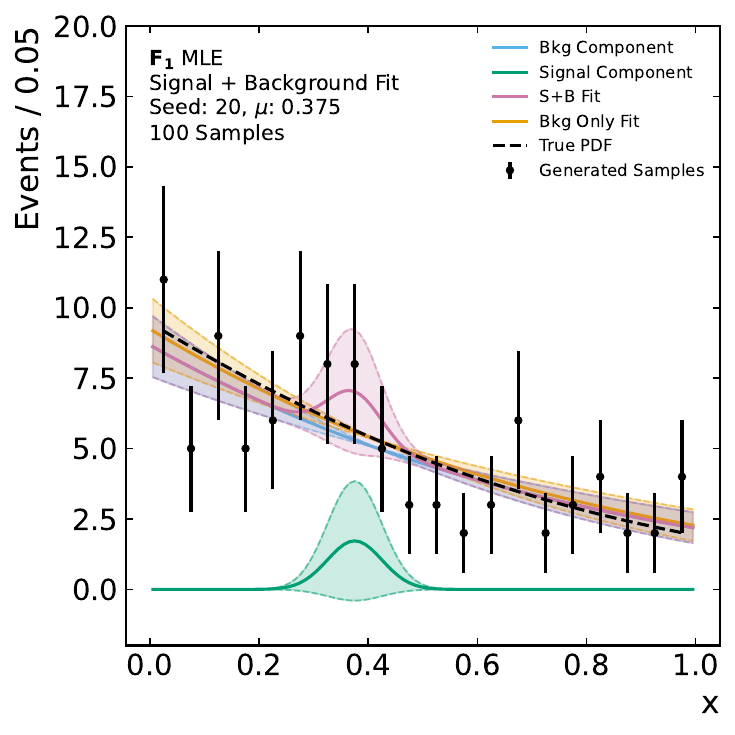}
\end{subfigure}
\hfill
\begin{subfigure}{0.32\linewidth}
     \includegraphics[width=\linewidth]{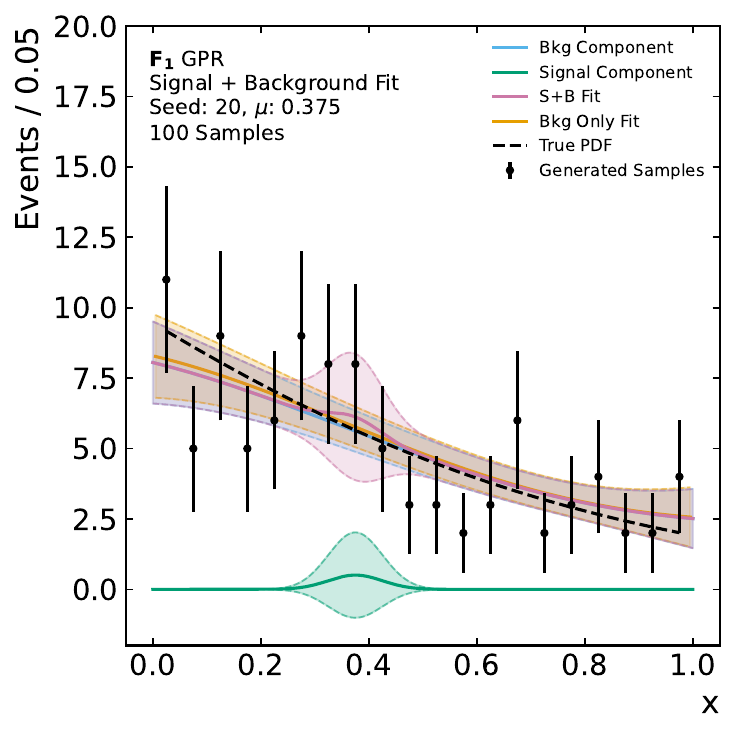}
\end{subfigure}
\caption{\label{fig:F1_SS_Example} A comparison of signal+background (pink) fit and its components (green and blue respectively) to background only (orange) in an $F_1$ dataset, showing for each method statistical fluctuations interpreted as signal. \rev{The signal component is scanned at $x = 0.375$, and the toy dataset contains 100 events.}}
\end{figure}

\begin{figure}[H]
\centering
\begin{subfigure}{0.32\linewidth}
     \includegraphics[width=\linewidth]{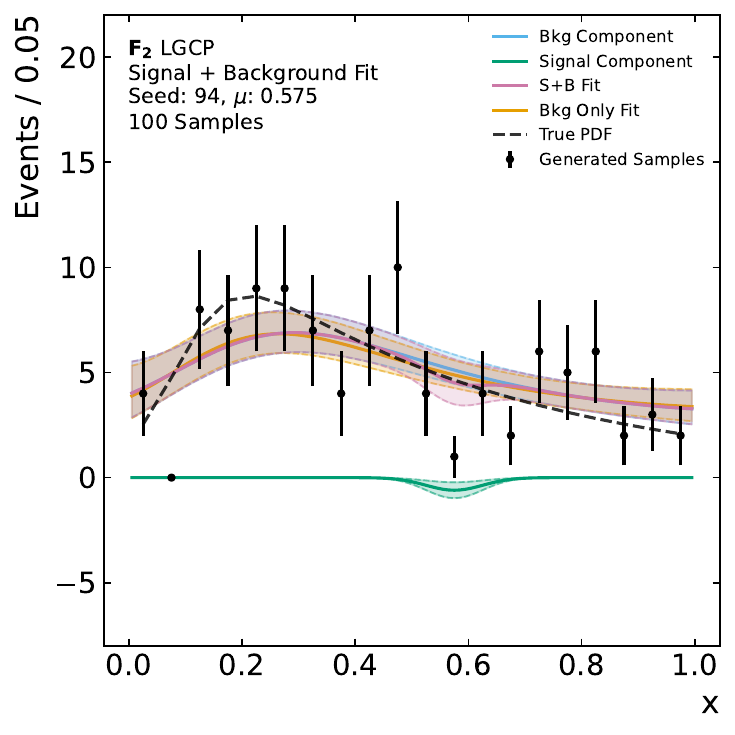}
\end{subfigure}
\hfill
\begin{subfigure}{0.32\linewidth}
     \includegraphics[width=\linewidth]{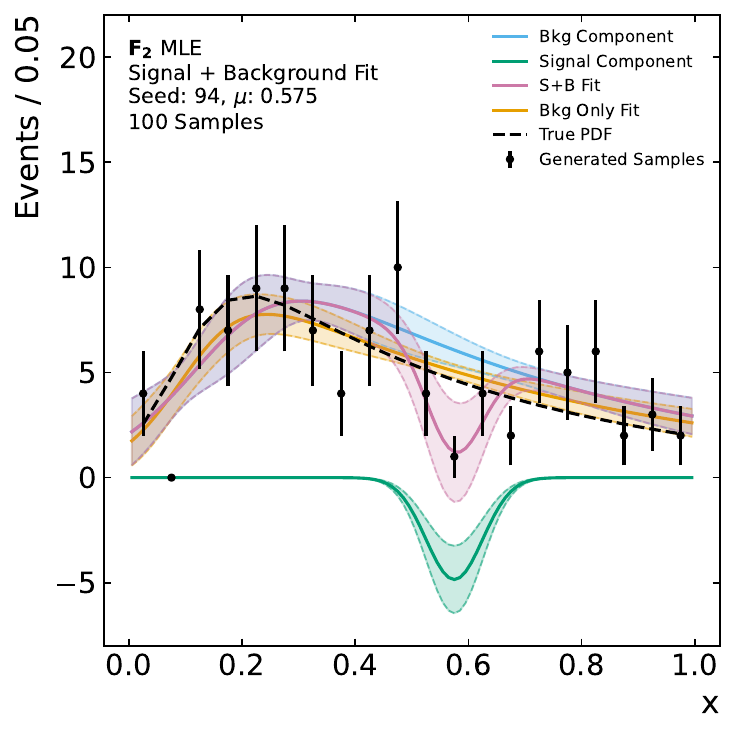}
\end{subfigure}
\hfill
\begin{subfigure}{0.32\linewidth}
     \includegraphics[width=\linewidth]{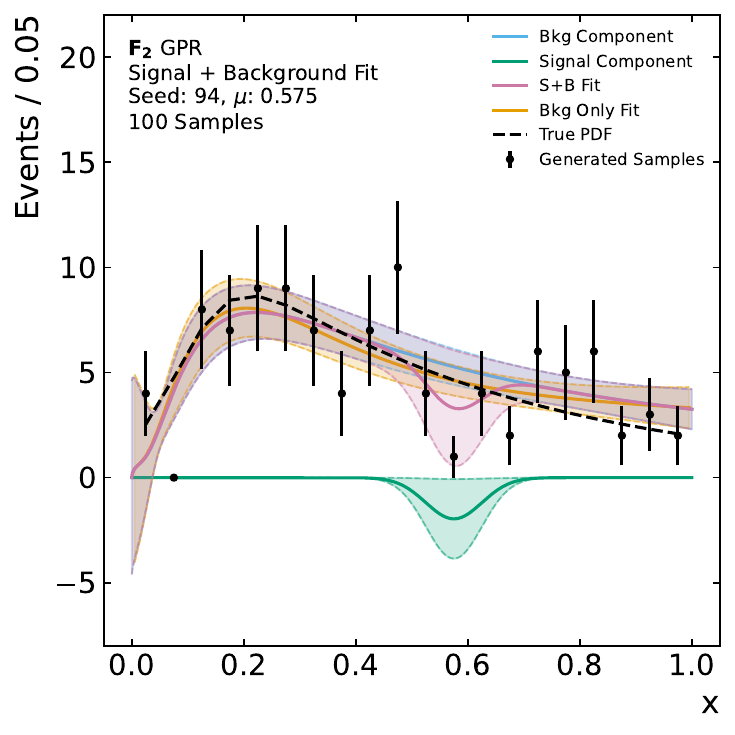}
\end{subfigure}
\caption{\label{fig:F2_SS_Example} A comparison of signal+background (pink) fit and its components (green and blue respectively) to background only (orange) in an $F_2$ dataset, showing for each method a turn on feature modeled as a signal. \rev{The signal component is scanned at $x = 0.575$, and the toy dataset contains 100 events.}}
\end{figure}

\begin{figure}[h]
\centering
\begin{subfigure}{0.32\linewidth}
     \includegraphics[width=\linewidth]{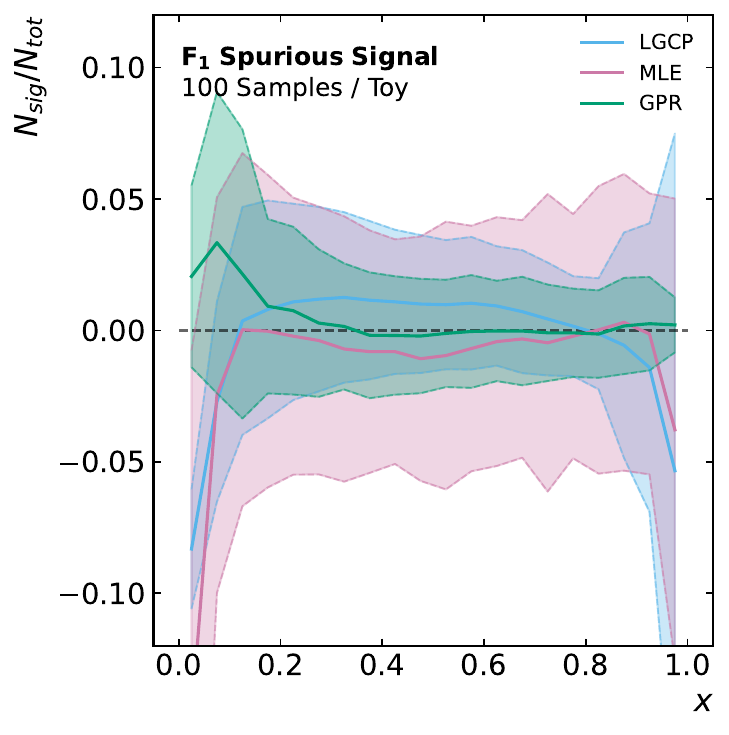}
\end{subfigure}
\hfill
\begin{subfigure}{0.32\linewidth}
     \includegraphics[width=\linewidth]{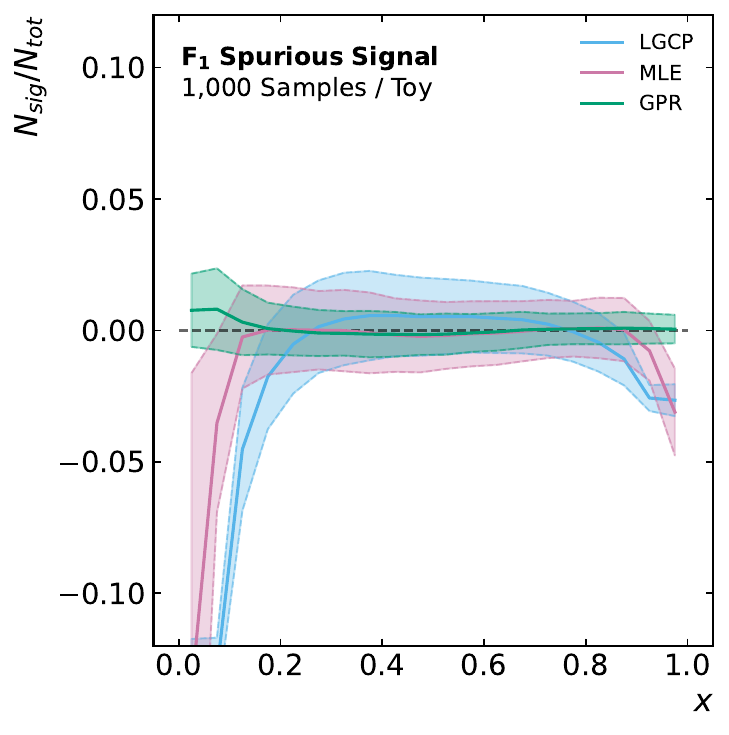}
\end{subfigure}
\hfill
\begin{subfigure}{0.32\linewidth}
     \includegraphics[width=\linewidth]{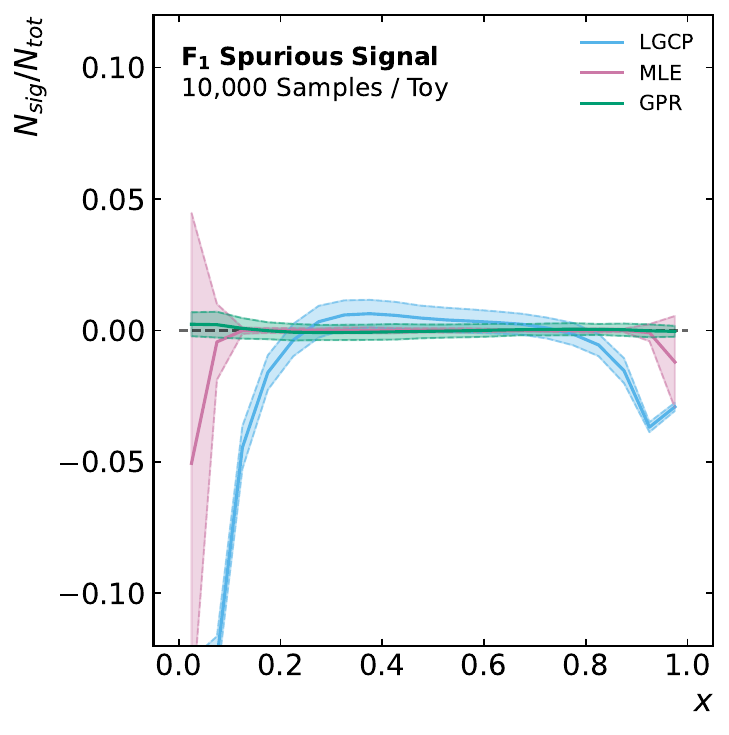}
\end{subfigure}
\caption{\label{fig:F1_SS} The average of the fitted spurious signal as a function of $x$ over 1000 $F_1$-generated toy datasets, for each fit strategy: (blue) LGCP, (green) GPR, (orange) MLE Estimate, and (pink) MLE. 
The spurious signal is shown normalized to the total number of samples in the fitted dataset: (a) 100, (b) 1000, and (c) $10,000$.
The shaded band indicates the $\pm 1 \sigma$ range of the relative fitted spurious signals over the 1000 toys. 
No signal contribution was injected into the toy datasets. }
\end{figure}

\begin{figure}[h]
\centering
\begin{subfigure}{0.32\linewidth}
     \includegraphics[width=\linewidth]{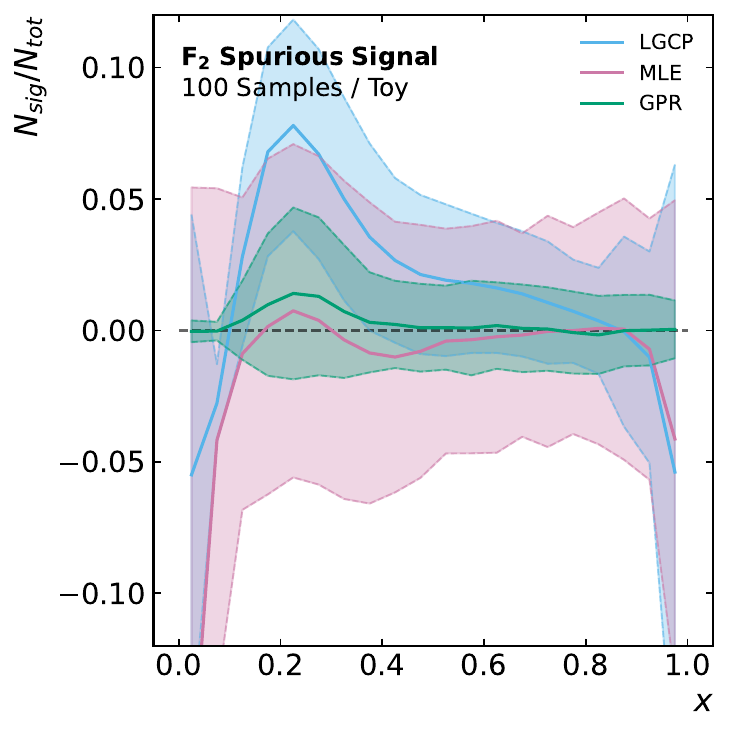}
\end{subfigure}
\hfill
\begin{subfigure}{0.32\linewidth}
     \includegraphics[width=\linewidth]{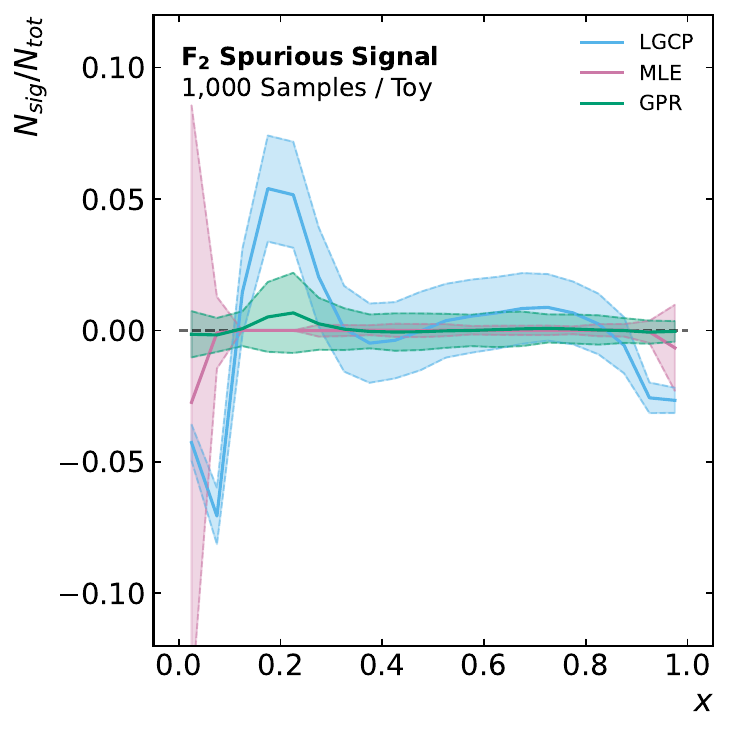}
\end{subfigure}
\hfill
\begin{subfigure}{0.32\linewidth}
     \includegraphics[width=\linewidth]{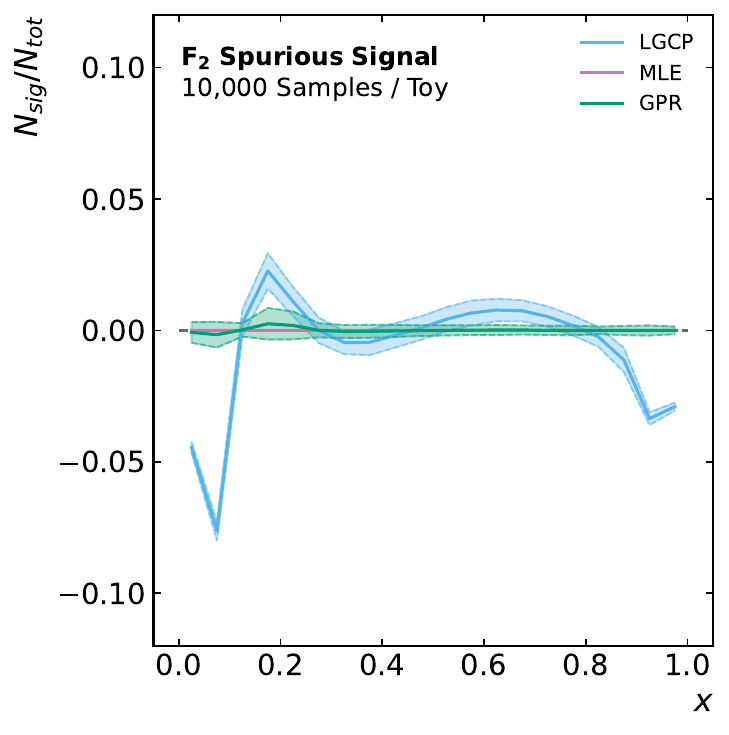}
\end{subfigure}\caption{\label{fig:F2_SS} The average of the fitted spurious signal as a function of $x$ over 1000 $F_2$-generated toy datasets, for each fit strategy: (blue) LGCP, (green) GPR, (orange) MLE Estimate, and (pink) MLE. 
The spurious signal is shown normalized to the total number of samples in the fitted dataset: (a) 100, (b) 1000, and (c) $10,000$.
The shaded band indicates the $\pm 1 \sigma$ range of the relative fitted spurious signals over the 1000 toys. 
No signal contribution was injected into the toy datasets. }
\end{figure}

For the MLE and LGCP fits, the signal is modeled using a Gaussian PDF with width of 0.05, while for the GP fits, the signal is modeled using the localized kernel described in Section~\ref{sec:SS} with the hyperparameter $t$ fixed to 0.05. 
For all three methods, the mean of the signal component of the fit (the $m$ hyperparameter of the GP signal kernel) is fixed in each fit, with mean values chosen scanning the full $x$ range. 
Examples of combined signal+background fits are performed using the LGCP, GP, and MLE methods. Examples for each function are shown in Figure~\ref{fig:F1_SS_Example} and Figure~\ref{fig:F2_SS_Example}. 
Figure~\ref{fig:F1_SS} shows the spurious signal, averaged across all 1000 toy datasets generated from $F_1$, for all three fit methods. 
The uncertainty bands show the $\pm 1 \sigma$ band of fitted spurious signal values for the 1000 toy dataset fits. 
Figure~\ref{fig:F2_SS} shows the same for $F_2$, respectively. 

\begin{figure}[hbt]
\centering
\begin{subfigure}{0.32\linewidth}
     \includegraphics[width=\linewidth]{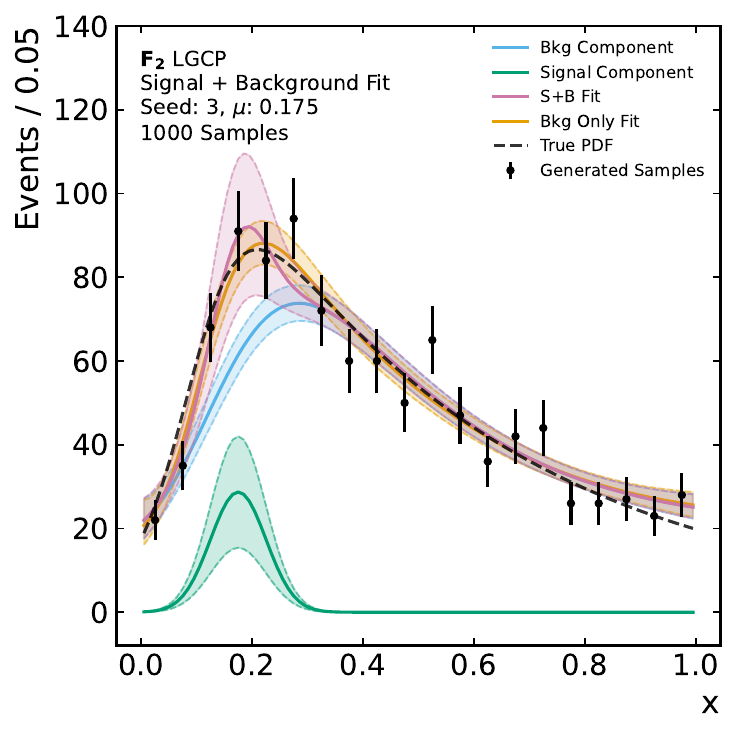}
\end{subfigure}
\hfill
\begin{subfigure}{0.32\linewidth}
     \includegraphics[width=\linewidth]{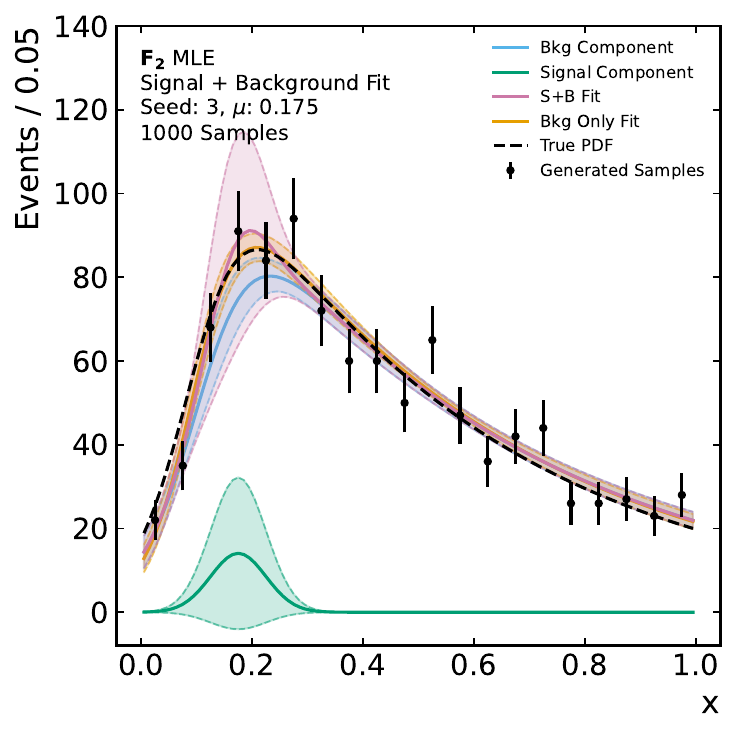}
\end{subfigure}
\hfill
\begin{subfigure}{0.32\linewidth}
     \includegraphics[width=\linewidth]{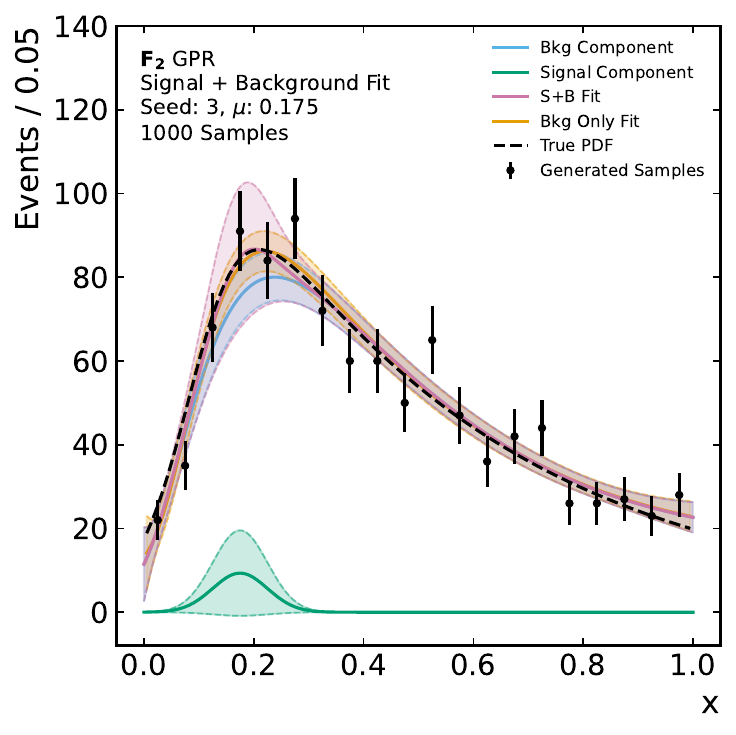}
\end{subfigure}
\caption{\label{fig:Failed_Fits} A comparison of signal+background (pink) fit and its components (green and blue respectively) to background only (orange) in an F2 dataset \rev{with 1000 events}, showing for each method a turn on feature modeled as a signal.}
\end{figure}

Near the middle of the $x$-range, the LGCP, GPR, and MLE generally produce a spurious signal corresponding to about 2\% or less of the total toy dataset statistics. 
The LGCP does demonstrate bias when fitting the turn-on feature in $F_2$, as shown in Figure~\ref{fig:Failed_Fits}, and it continues to demonstrate biases near the edges of the $x$-range, as seen in the pull results. 
The MLE also demonstrates anomalously large spurious signal values near the low-edge of the $x$ range for both $F_1$ and $F_2$, but unlike LGCP, isn't impacted by the turn on feature. 
The fitted spurious signal from the GPR is generally observed to be compatible with 0 within uncertainties, and also not impacted by the turn on feature.

\FloatBarrier
\subsection{Injection Tests}
The next stage is to perform the spurious signal test after injecting a signal \rev{(modeled here as a localized Gaussian distribution)} to show that the test is indeed sensitive to a real resonance in the data set.

\begin{figure}[hbt]
\centering
\begin{subfigure}{0.32\linewidth}
    \includegraphics[width=\linewidth]{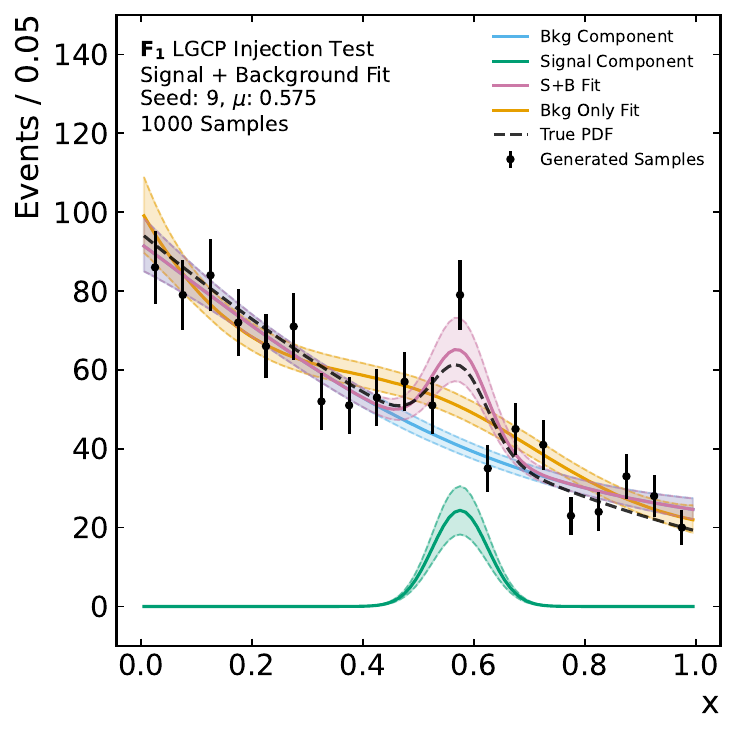}
\end{subfigure}
\hfill
\begin{subfigure}{0.32\linewidth}
    \includegraphics[width=\linewidth]{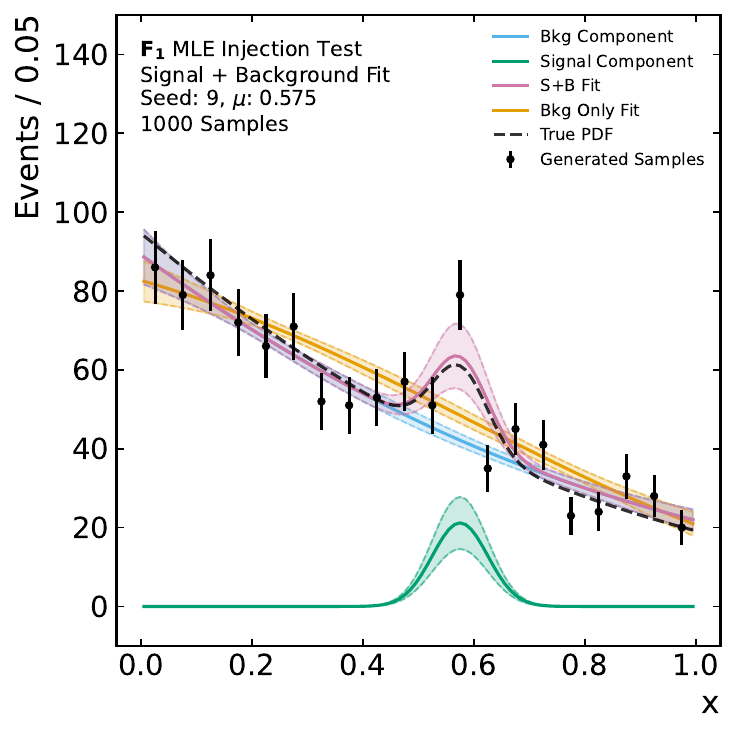}
\end{subfigure}
\hfill
\begin{subfigure}{0.32\linewidth}
    \includegraphics[width=\linewidth]{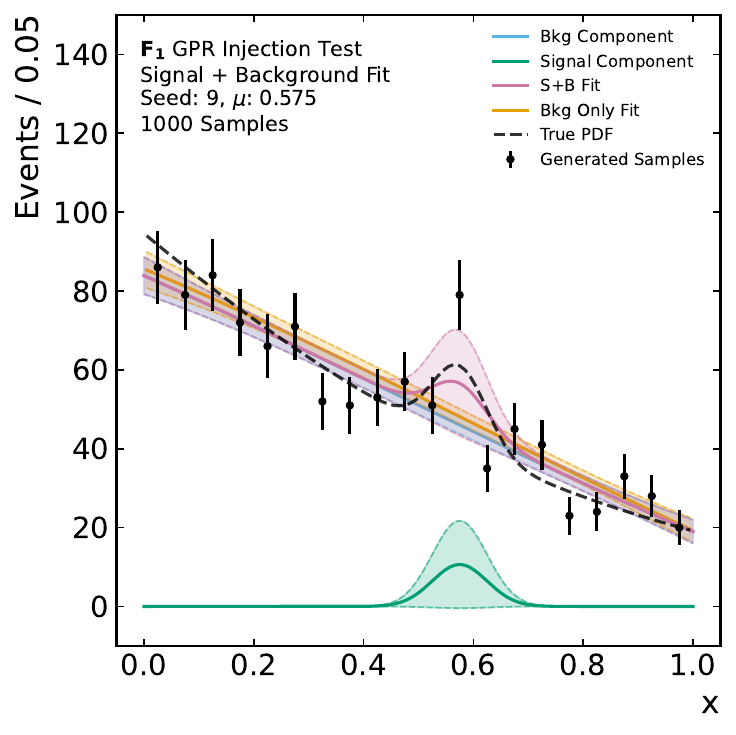}
\end{subfigure}

\caption{\label{fig:$F_1$_Inj_Example} A comparison of signal+background (pink) fit and its components (green and blue respectively) to background only (orange)  in an $F_1$ dataset after injection, showing for each method the sensitivity to signal detection. }
\end{figure}

\begin{figure}[hbt]
\centering
\begin{subfigure}{0.32\linewidth}
    \includegraphics[width=\linewidth]{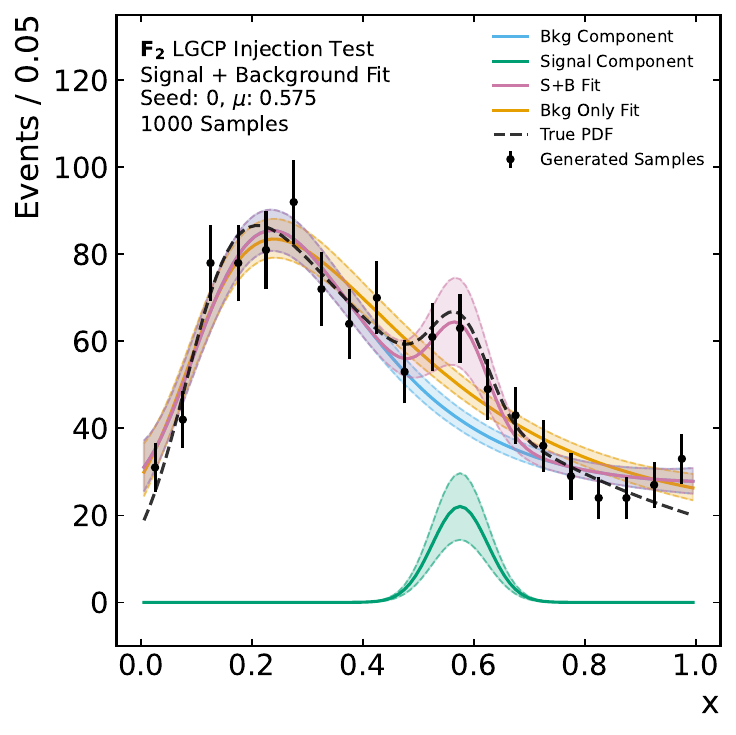}
\end{subfigure}
\hfill
\begin{subfigure}{0.32\linewidth}
    \includegraphics[width=\linewidth]{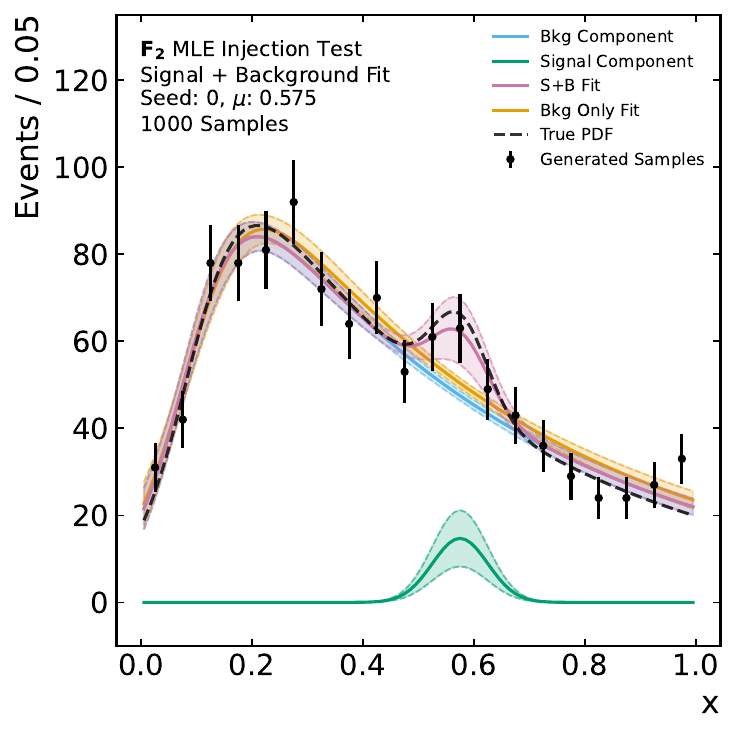}
\end{subfigure}
\hfill
\begin{subfigure}{0.32\linewidth}
    \includegraphics[width=\linewidth]{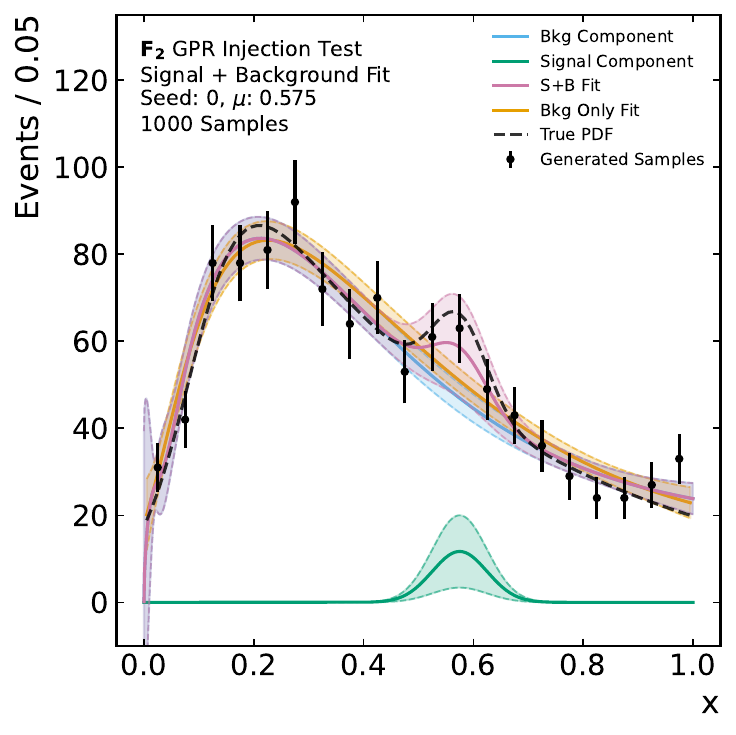}
\end{subfigure}

\caption{\label{fig:$F_2$_Inj_Example} A comparison of signal+background (pink) fit and its components (green and blue respectively) to background only (orange)  in an $F_2$ dataset after injection, showing for each method the sensitivity to signal detection. }
\end{figure}

First, a linear test is performed: a Gaussian signal bump with $\sigma = 0.05$ is injected at $x = 0.575$ for each data set, with a linear increase from 1\% up to 15\% of the total data set size, in 1\% increments. For example, a toy data set with 1000 events with \rev{5\%} injected signal will have now \rev{1050} total events. 

Examples of LGCP, GPR, and MLE combined signal$+$background fits of toys generated from functions $F_1$ and $F_2$ are shown in Figure~\ref{fig:$F_1$_Inj_Example} and \rev{Figure}~\ref{fig:$F_2$_Inj_Example}, respectively. 
The example fits show an injected signal at \rev{$x=0.575$}, near the middle of the $x$-range, and the toys each contain $1000$ background events and $50$ injected signal events. 
The example fits show reasonable behavior for all methods, with the injected signal generally being captured as such by the signal component of each fit. 

\begin{figure}[H]
\centering
\begin{subfigure}{0.32\linewidth}
    \includegraphics[width=\linewidth]{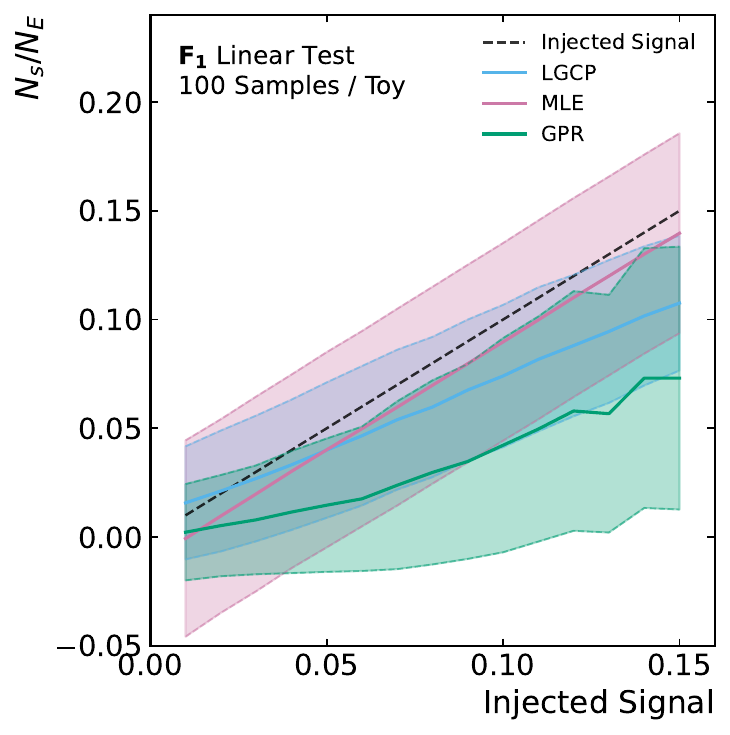}

\end{subfigure}
\hfill
\begin{subfigure}{0.32\linewidth}
    \includegraphics[width=\linewidth]{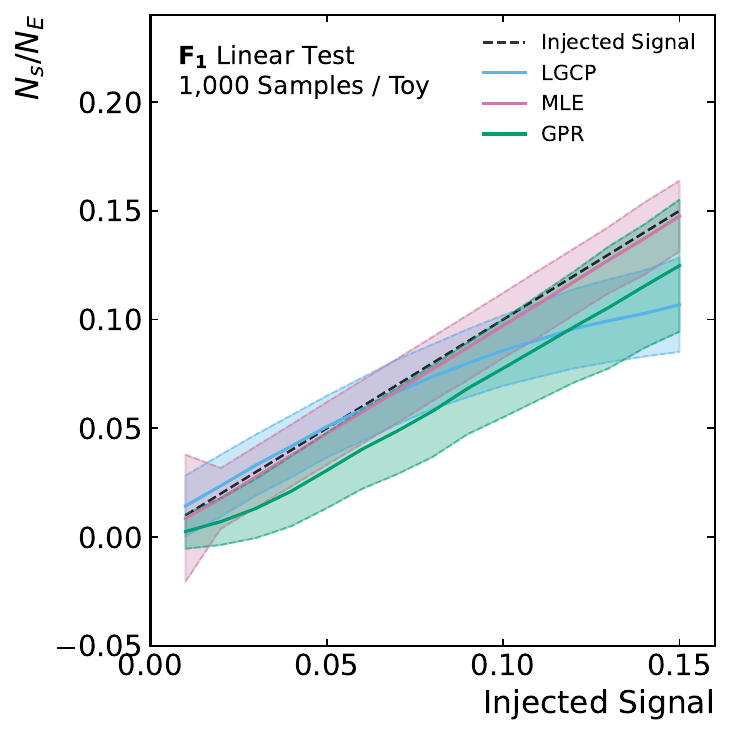}
\end{subfigure}
\hfill
\begin{subfigure}{0.32\linewidth}
    \includegraphics[width=\linewidth]{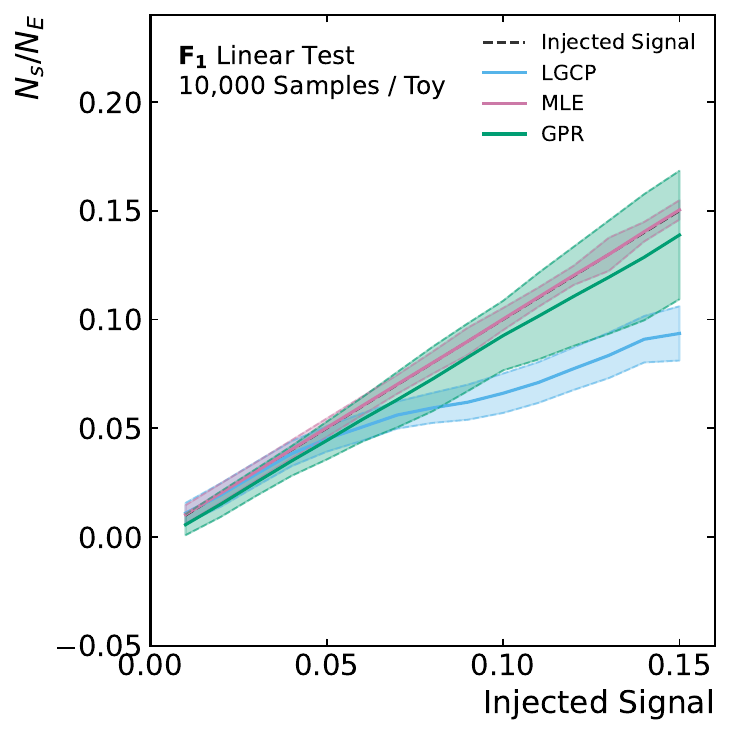}
\end{subfigure}

\caption{\label{fig:F1_Linear} The average of the fitted spurious signal for each of the three fit strategies over 1000 $F_1$-generated toy datasets as a function of the injected signal yield. 
The spurious signal is shown normalized to the total number of samples in the fitted dataset: (a) 100, (b) 1000, and (c) $10,000$.
The signal was injected at $x=0.575$. }
\end{figure}

\begin{figure}[H]
\centering
\begin{subfigure}{0.32\linewidth}
    \includegraphics[width=\linewidth]{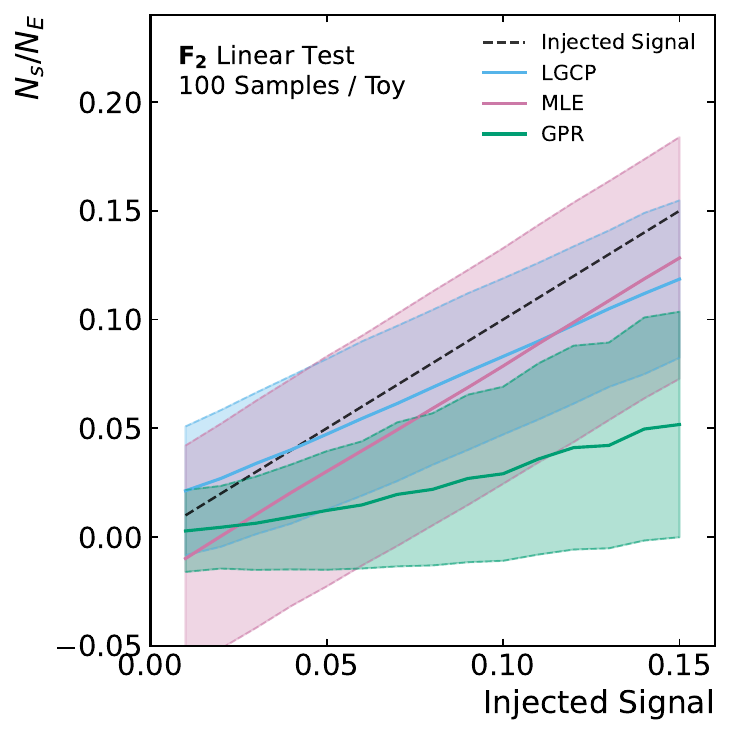}
\end{subfigure}
\hfill
\begin{subfigure}{0.32\linewidth}
    \includegraphics[width=\linewidth]{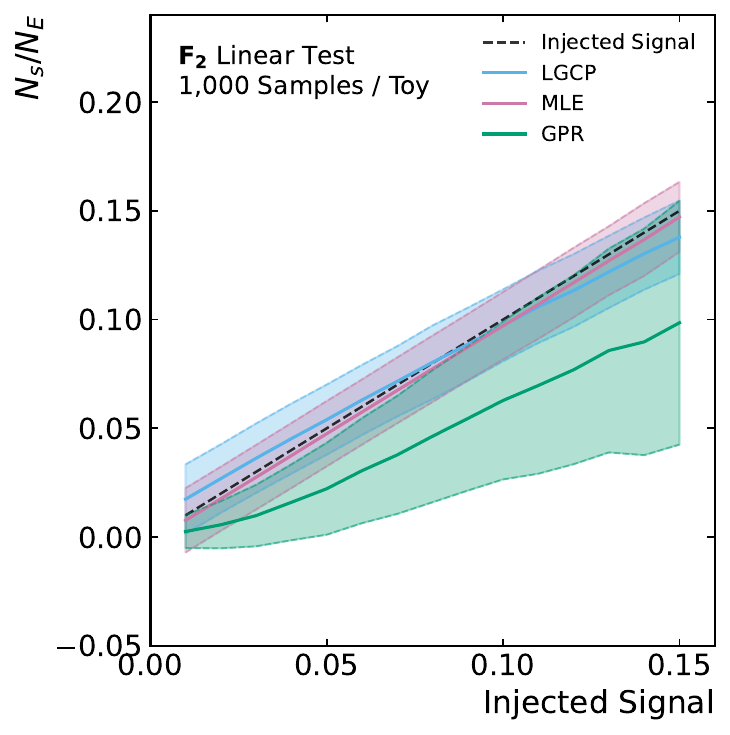}
\end{subfigure}
\hfill
\begin{subfigure}{0.32\linewidth}
    \includegraphics[width=\linewidth]{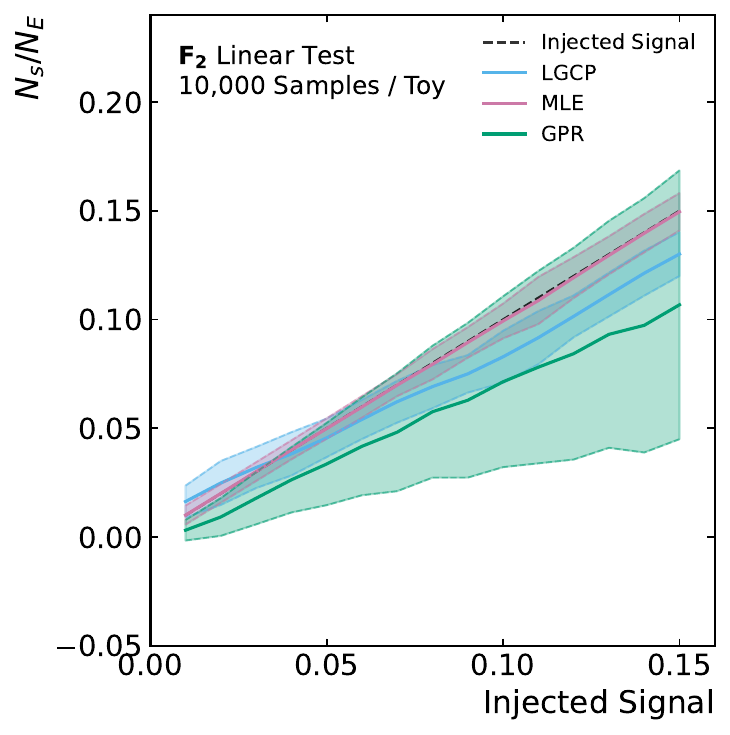}
\end{subfigure}

\caption{\label{fig:F2_Linear} The average of the fitted spurious signal for each of the three fit strategies over 1000 $F_2$-generated toy datasets as a function of the injected signal yield. 
The spurious signal is shown normalized to the total number of samples in the fitted dataset: (a) 100, (b) 1000, and (c) $10,000$.
The signal was injected at $x=0.575$.}
\end{figure}

Figure~\ref{fig:F1_Linear} and \rev{Figure}~\ref{fig:F2_Linear} summarize the results of the signal injection tests and show the average fitted signal yield for all 1000 toy datasets by each fitting method, for injected signal yields between $1-15\%$ of the total background yield. 
As can be observed, the LGCP generally capturs the injected signal well in both functions up to around 5\% signal yield, regardless of the overally statistics of the toy datasets. 
Above 5\% signal yield, the LGCP starts to heavily underestimate the amount of signal injected, with results getting worse as statistics increase. 
The MLE shows an almost identical trend to the actual injection, with results improving with increasing statistics. 
The GPR significantly underestimates the magnitude of the injected signal in the lower-statistics case (100 samples per toy-dataset). 
Other than for the highest-statistics $F_1$ toys, the GPR captures a smaller frasction of the injected signal than the LGCP method. 
While increasing the toy dataset statistics improves the mean of the fitted GPR signal yield, the standard deviation of the fitted signal yield over the toy datasets remains large compared with that of the LCGP or MLE methods.

Following the linear injection tests, further tests were performed with a fixed $5\%$ injected signal yield, but with different $x$ locations for the injected signal: $x = 0.075, 0.575, 0.925$. These tests will also include the edge effects in the GPR or LCGP fits. 

\begin{figure}[hbt]
\centering
\begin{subfigure}{0.32\linewidth}
    \includegraphics[width=\linewidth]{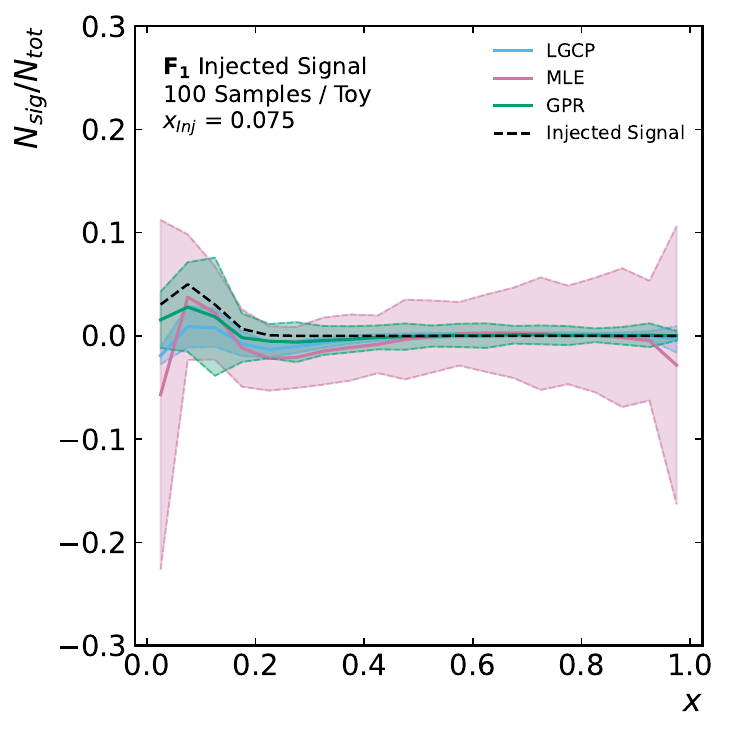}
\end{subfigure}
\hfill
\begin{subfigure}{0.32\linewidth}
    \includegraphics[width=\linewidth]{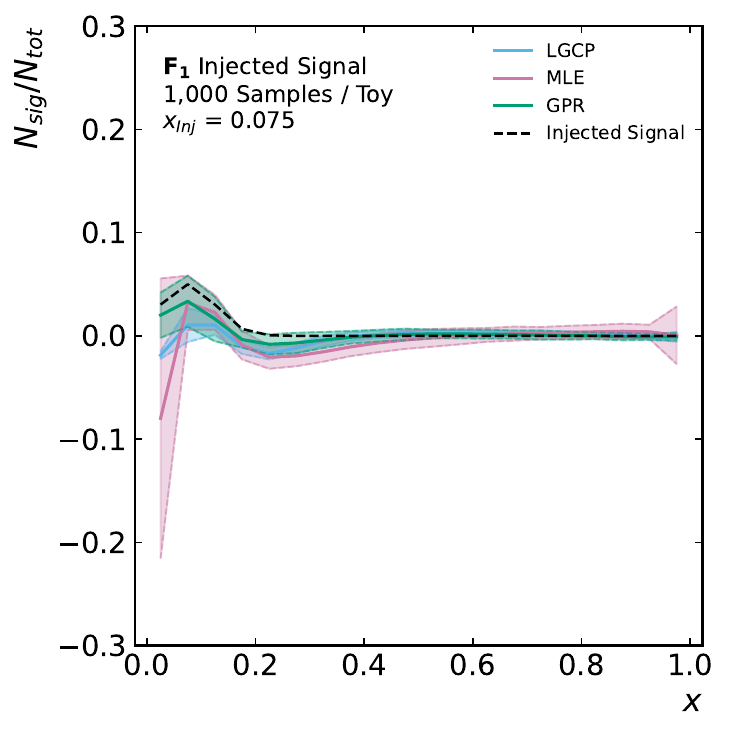}
\end{subfigure}
\hfill
\begin{subfigure}{0.32\linewidth}
    \includegraphics[width=\linewidth]{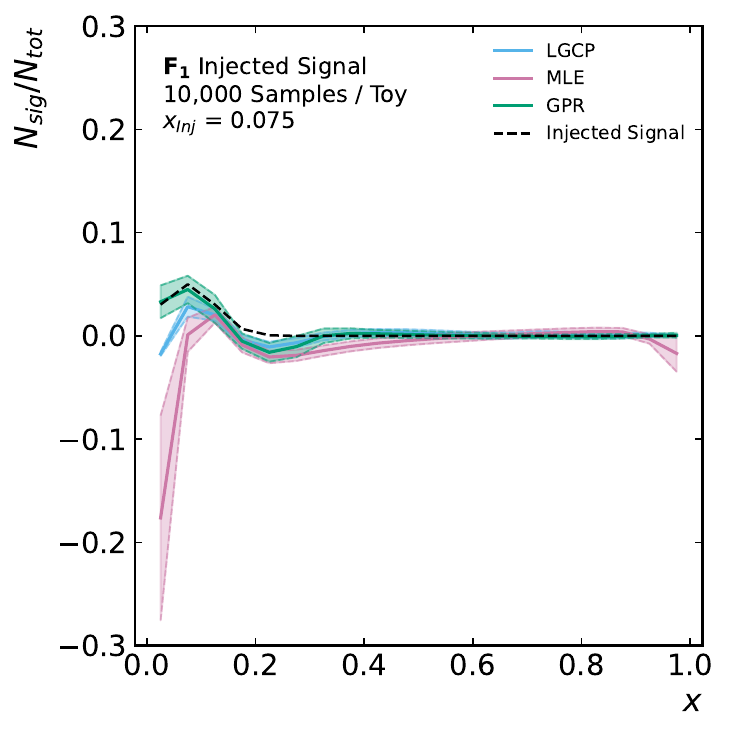}
\end{subfigure}

\vspace{0.5em}

\begin{subfigure}{0.32\linewidth}
    \includegraphics[width=\linewidth]{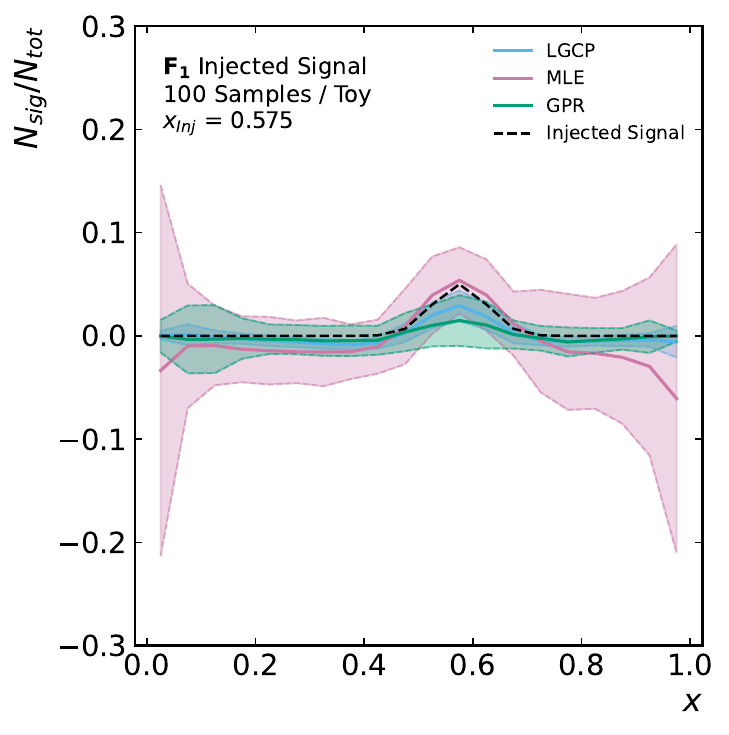}
\end{subfigure}
\hfill
\begin{subfigure}{0.32\linewidth}
    \includegraphics[width=\linewidth]{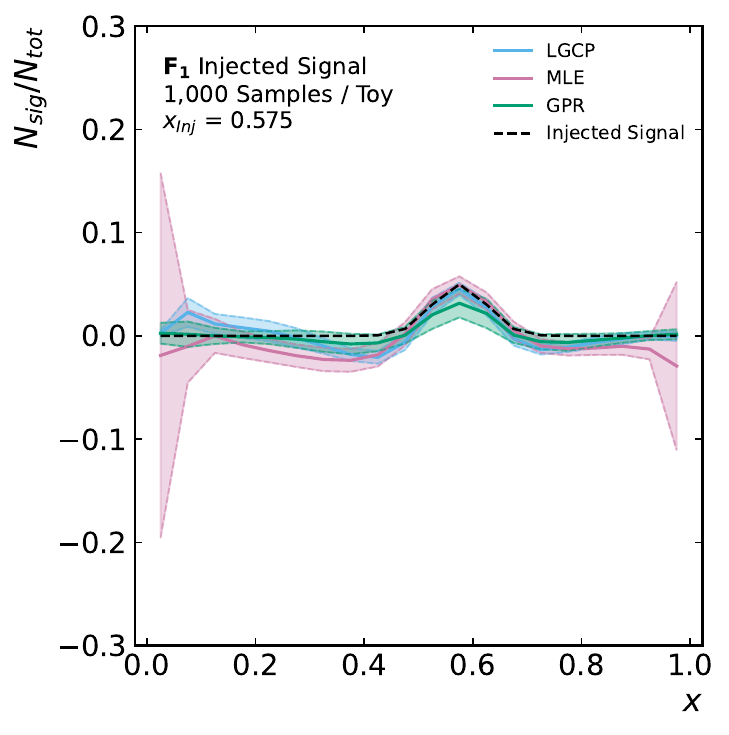}
\end{subfigure}
\hfill
\begin{subfigure}{0.32\linewidth}
    \includegraphics[width=\linewidth]{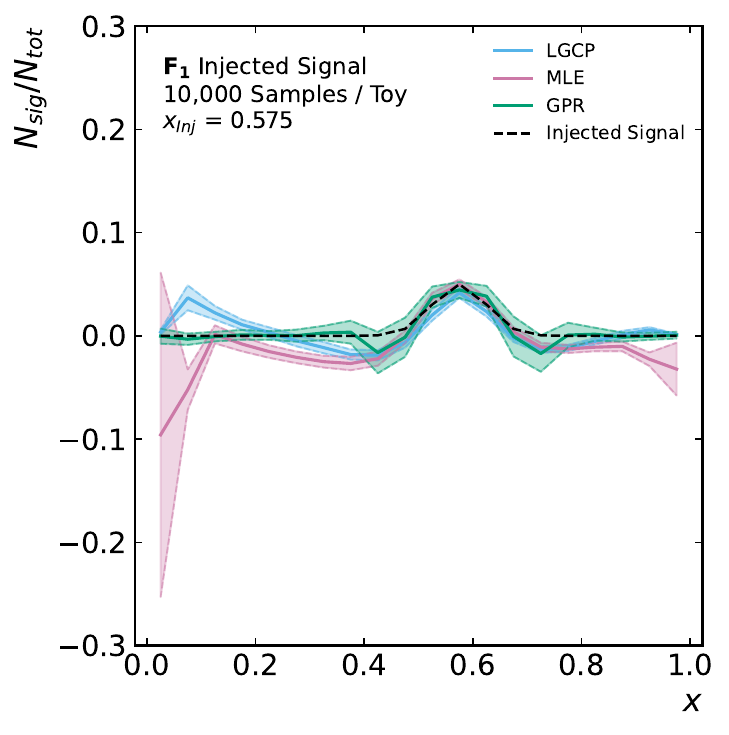}
\end{subfigure}

\vspace{0.5em}

\begin{subfigure}{0.32\linewidth}
    \includegraphics[width=\linewidth]{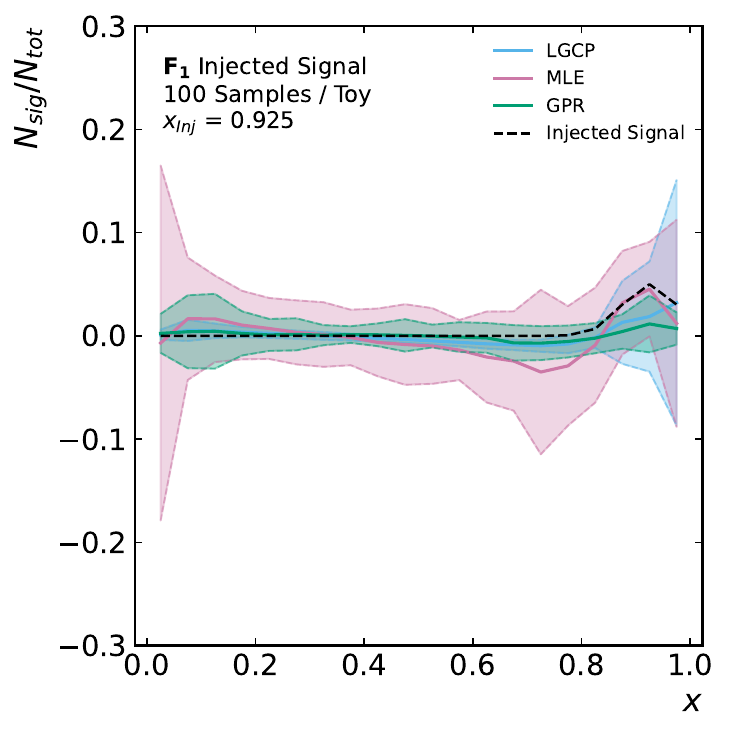}
\end{subfigure}
\hfill
\begin{subfigure}{0.32\linewidth}
    \includegraphics[width=\linewidth]{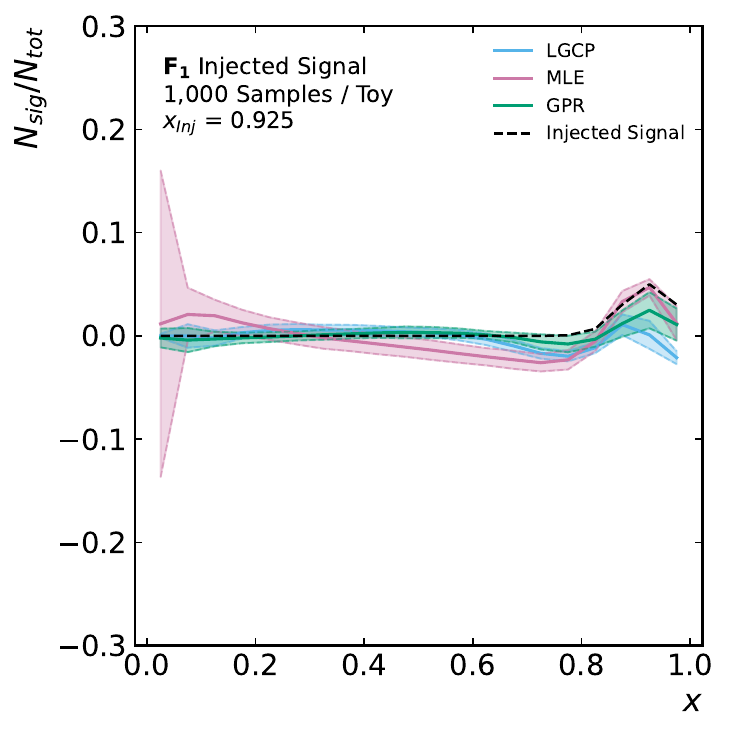}
\end{subfigure}
\hfill
\begin{subfigure}{0.32\linewidth}
    \includegraphics[width=\linewidth]{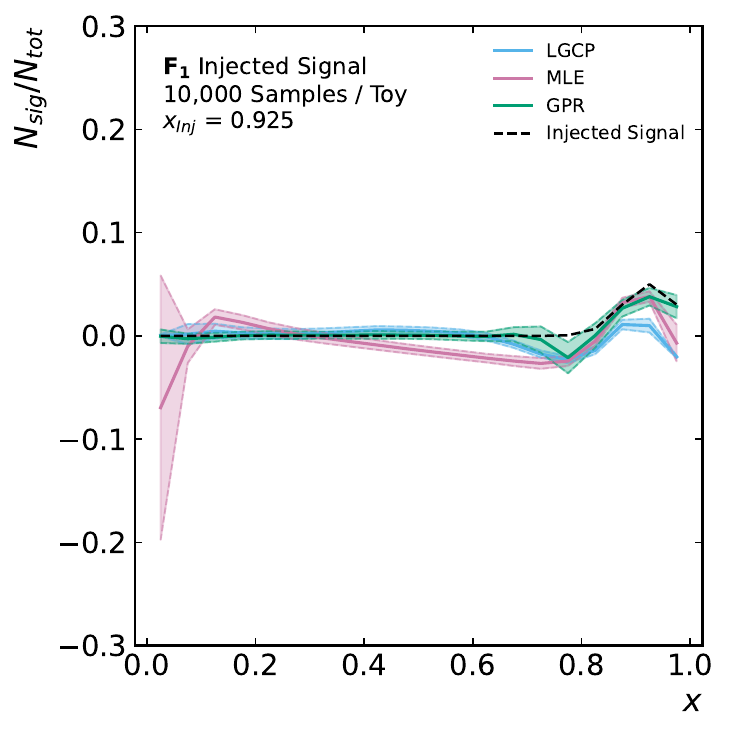}
\end{subfigure}
\caption{\label{fig:F1_Inj} The difference in the average measured spurious signal yield over the 1000 $F_1$-generated toy datasets, with and without an injected signal, at: (top row) $x=0.075$, (middle row) $x=0.575$, and (bottom row) $x=0.925$ for the three strategies. The difference is normalized by the number of samples in the toy datasets: (left column) 100 events, (middle column) $1000$ events, and (right column) $10000$ events. The shaded band shows the $\pm 1 \sigma$ spread of the difference in fitted spurious signal yield (with versus without an injected signal) over the 1000 toys. The dashed line shows the expected difference when the fit strategy successfully captures the injected signal with no impact on the background fit: $5\%$ at the site of the injected signal and $0$ for all other points in $x$.}
\end{figure}

\begin{figure}[hbt]
\centering
\begin{subfigure}{0.32\linewidth}
    \includegraphics[width=\linewidth]{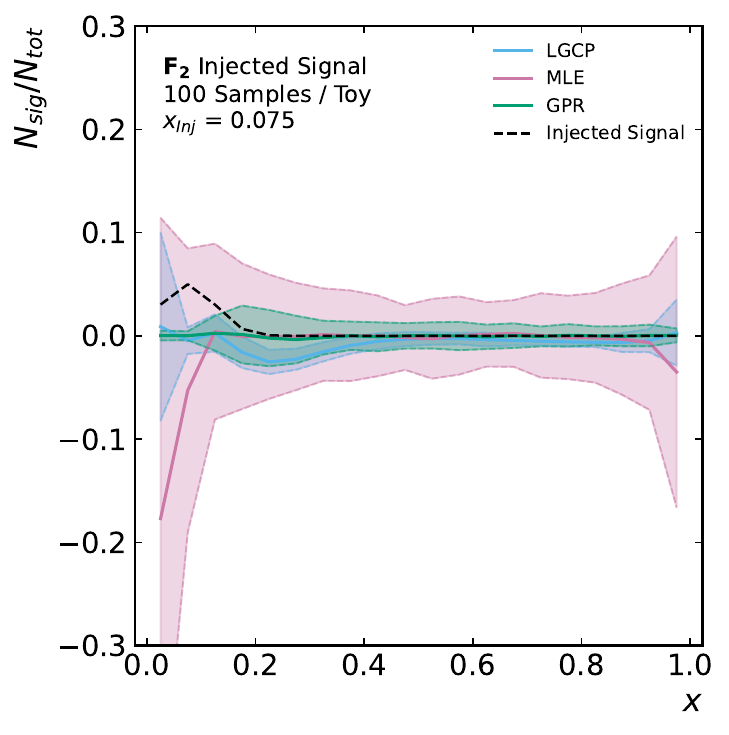}
\end{subfigure}
\hfill
\begin{subfigure}{0.32\linewidth}
    \includegraphics[width=\linewidth]{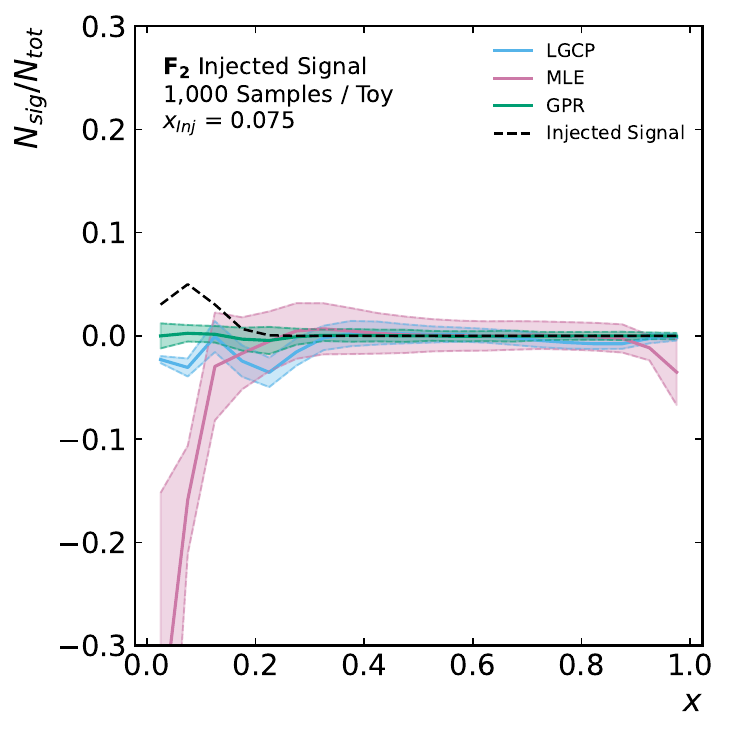}
\end{subfigure}
\hfill
\begin{subfigure}{0.32\linewidth}
    \includegraphics[width=\linewidth]{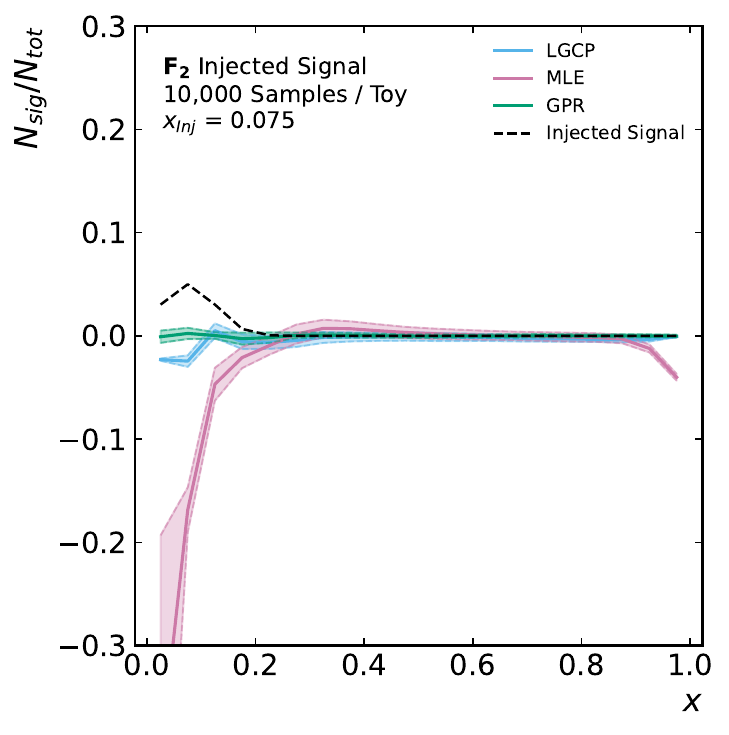}
\end{subfigure}

\vspace{0.5em}

\begin{subfigure}{0.32\linewidth}
    \includegraphics[width=\linewidth]{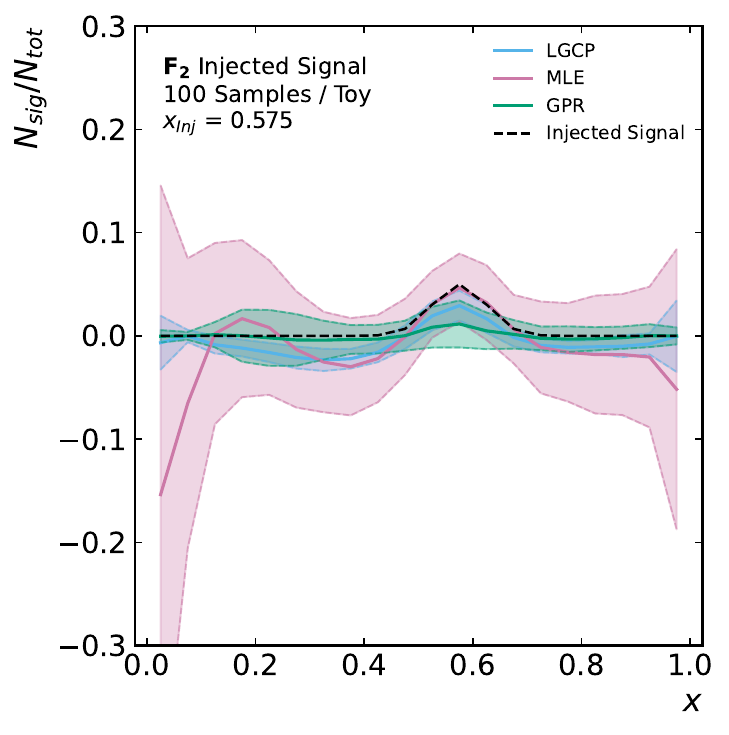}
\end{subfigure}
\hfill
\begin{subfigure}{0.32\linewidth}
    \includegraphics[width=\linewidth]{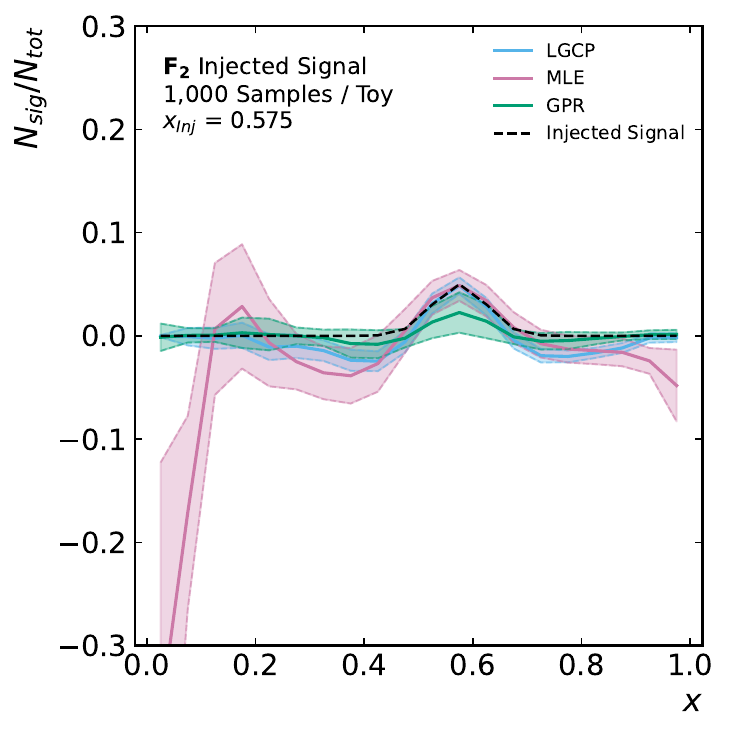}
\end{subfigure}
\hfill
\begin{subfigure}{0.32\linewidth}
    \includegraphics[width=\linewidth]{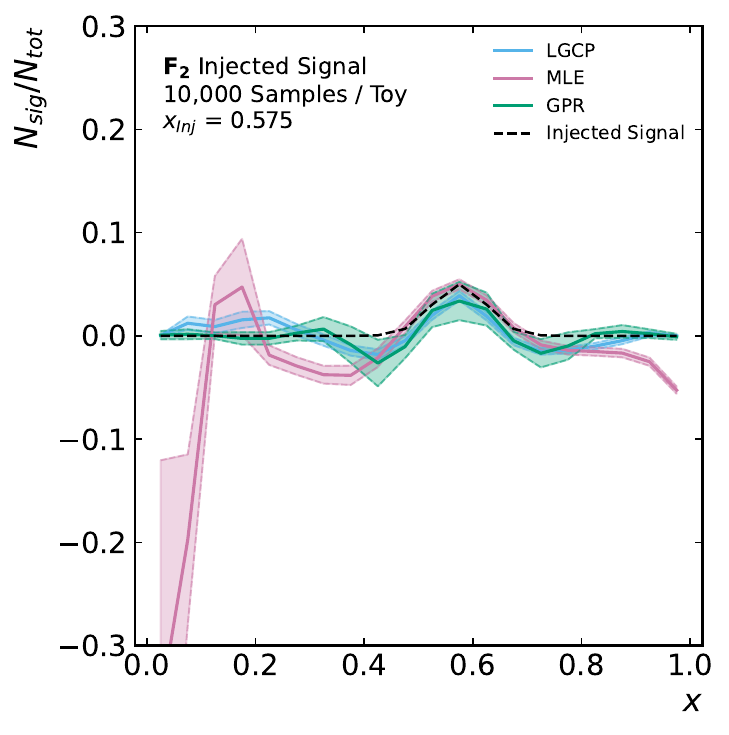}
\end{subfigure}

\vspace{0.5em}

\begin{subfigure}{0.32\linewidth}
    \includegraphics[width=\linewidth]{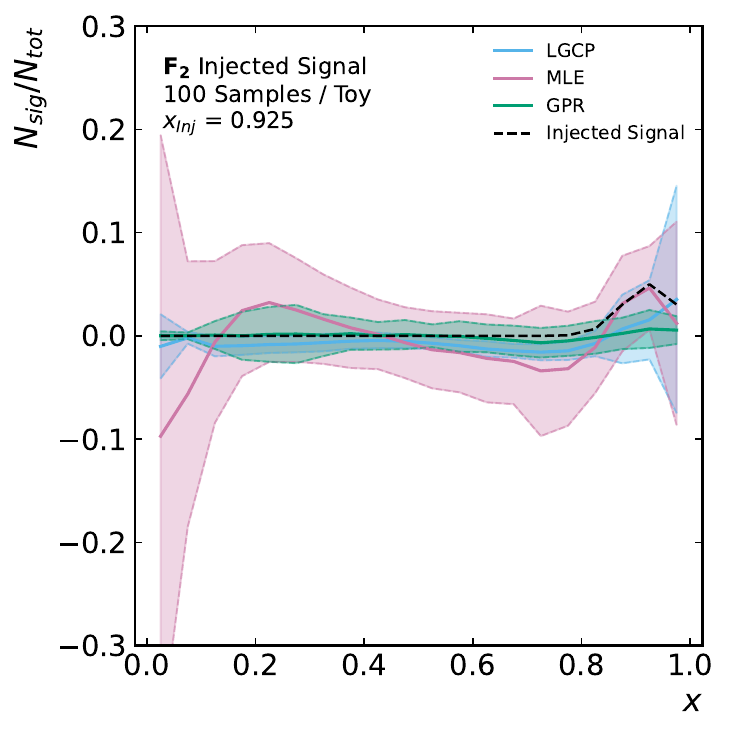}
\end{subfigure}
\hfill
\begin{subfigure}{0.32\linewidth}
    \includegraphics[width=\linewidth]{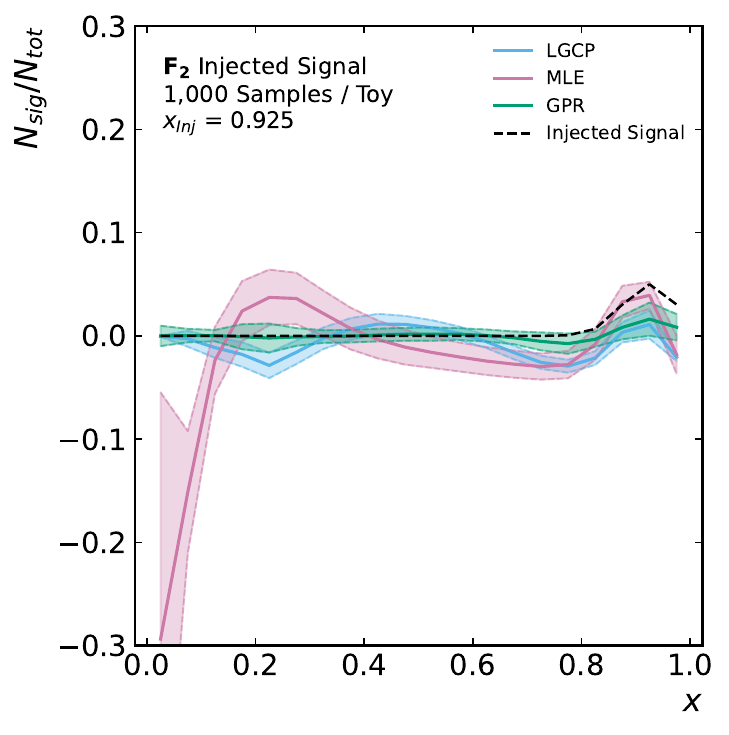}
\end{subfigure}
\hfill
\begin{subfigure}{0.32\linewidth}
    \includegraphics[width=\linewidth]{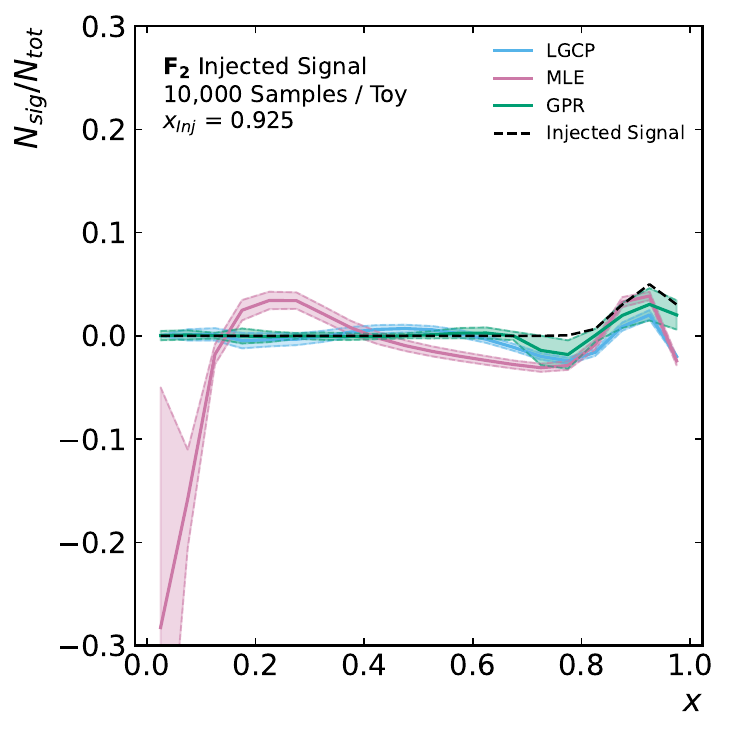}
\end{subfigure}
\caption{\label{fig:F2_Inj} The difference in the average measured spurious signal yield over the 1000 $F_2$-generated toy datasets, with and without an injected signal, at: (top row) $x=0.075$, (middle row) $x=0.575$, and (bottom row) $x=0.925$ for the three strategies. The difference is normalized by the number of samples in the toy datasets: (left column) 100 events, (middle column) $1000$ events, and (right column) $10000$ events. The shaded band shows the $\pm 1 \sigma$ spread of the difference in fitted spurious signal yield (with versus without an injected signal) over the 1000 toys. The dashed line shows the expected difference when the fit strategy successfully captures the injected signal with no impact on the background fit: $5\%$ at the site of the injected signal and $0$ for all other points in $x$.}
\end{figure}

Figure~\ref{fig:F1_Inj} and \rev{Figure}~\ref{fig:F2_Inj} show the spurious signal test for each working point, for $F_1$ and $F_2$ respectively. 
For $F_1$ the LGCP effectively captures the signal injection in the middle, improving with increasing statistics, while the edge injections are fitted poorly. 
The MLE perform well in all cases, but also shows edge effects for the low-$x$ range. 
The ability of the GP to effectively capture the injected signal is relatively insensitive to the location of the injection, but for low statistics toy datasets, it generally assigns most of the signal yield as background events (as seen by the systematic undershoot of the GPR fitted signal yield).  
For $F_2$, the results are quite different owing to the turn on having a similar shape to a resonance. 
Examples of failed LGCP fits to toy datasets generated from $F_2$ can be seen in Figure~\ref{fig:Failed_Fits}, in which either the turn-on feature of the underlying function or a statistical fluctuation is spuriously fitted as a signal. 
The LGCP continues to fail in the edges, but the signal injected in the middle of the $x$-range is again effectively fitted. The MLE performs similarly to LGCP for the low $x$-range and the center, and it is also able to fit the signal injected at the high $x$-range. 
However, it also was observed to spuriously fit the turn-on as signal, similar to the LGCP fit examples shown in Figure~\ref{fig:Failed_Fits}. 
The GP generally avoids fitting the turn-on as a signal, unlike the LGCP and MLE, but it also fails to fit the injected signal as such for low- and mid-statistics toy datasets. 


\FloatBarrier

\section{Summary}

A novel application of Log Gaussian Cox Processes is applied to fitting one-dimensional smooth background distributions characteristic of many bump-hunt style analyses in LHC experiments. 
The LGCP was compared to the standard unbinned MLE fit and another gaussian process based method, binned GPR, using two different functional forms with three different levels of event statistics: 100, 1000, and $10,000$ events.

For the background only estimation, both LGCP and GPR performed well based on the pull plots, showcasing the option for automated background modeling without making assumptions on the functional form of the underlying PDF. 
For the Spurious Signal test, the GPR method showed stronger resilience in ignoring statistical fluctuations relative to both LGCP and MLE methods, but on the other hand wasn't sensitive enough to capture an actual signal injected in the injection tests. Unlike the GPR, the LGCP performed well in capturing injected signal of up to 5\% from the total amount of events.
Based on these results, assuming there are wide enough sidebands to avoid edge-effects, the LGCP method can be used for both background and signal+background modeling. Using LGCP can make these sections of future analyses faster and more efficient, without sacrificing accuracy. The GPR on the other hand doesn't capture injected signals as well, and seems to be ideal for smoothing existing background templates, similarly to the use case in \cite{diphoton}, before being fit with a different method.

\section*{Acknowledgments}
YF and LB are supported by an ERC STGgrant, grant No.945878. 
MK is supported by the US Department of Energy (DOE) under grant DE-AC02-76SF00515. 
During her contribution to this work, PJ was supported by a US Department of Energy (DOE) Summer Undergraduate Laboratory Internship (SULI) position hosted at SLAC. 
During her time at SLAC, RH was supported by the US Department of Energy (DOE) under grant DE-AC02-76SF00515.

\bibliographystyle{unsrturl} 
\bibliography{references}

@article{LGCP,
author = {Møller, J. and Syversveen, A. R. and Waagepetersen, R. P.},
title = "{Log Gaussian Cox Processes}",
journal = "{Scandinavian Journal of Statistics}",
volume = {25},
number = {3},
pages = {451-482},
doi = {https://doi.org/10.1111/1467-9469.00115},
url = {https://onlinelibrary.wiley.com/doi/abs/10.1111/1467-9469.00115},
year = {1998}
}

@misc{charm_higgs,
      title="{Search for the associated production of charm quarks and a Higgs boson decaying into a photon pair with the ATLAS detector}", 
      author="{ATLAS Collaboration}",
      year={2024},
      eprint={2407.15550},
      archivePrefix={arXiv},
      primaryClass={hep-ex},
      url={https://arxiv.org/abs/2407.15550}, 
}

@ARTICLE{MCMC,
       author = {{Hastings}, W.~K.},
        title = "{Monte Carlo Sampling Methods using Markov Chains and their Applications}",
      journal = {Biometrika},
         year = 1970,
        month = apr,
       volume = {57},
       number = {1},
        pages = {97-109},
          doi = {10.1093/biomet/57.1.97},
       adsurl = {https://ui.adsabs.harvard.edu/abs/1970Bimka..57...97H}
}

@techreport{ATL-PHYS-PUB-2020-028,
      author = "{ATLAS Collaboration}",
      title         = "{Recommendations for the Modeling of Smooth Backgrounds}",
      institution   = "CERN",
      reportNumber  = "ATL-PHYS-PUB-2020-028",
      address       = "Geneva",
      year          = "2020",
      url           = "https://cds.cern.ch/record/2743717",
      note          = "All figures including auxiliary figures are available at
                       https://atlas.web.cern.ch/Atlas/GROUPS/PHYSICS/PUBNOTES/ATL-PHYS-PUB-2020-028",
}

@techreport{ATL-PHYS-PUB-2022-045,
      author = "{ATLAS Collaboration}",
      title         = "{Active Learning reinterpretation of an ATLAS Dark Matter
                       search constraining a model of a dark Higgs boson decaying
                       to two b-quarks}",
      institution   = "CERN",
      reportNumber  = "ATL-PHYS-PUB-2022-045",
      address       = "Geneva",
      year          = "2022",
      url           = "https://cds.cern.ch/record/2839789",
      note          = "All figures including auxiliary figures are available at
                       https://atlas.web.cern.ch/Atlas/GROUPS/PHYSICS/PUBNOTES/ATL-PHYS-PUB-2022-045",
}

@article{Dauncey_2015,
doi = {10.1088/1748-0221/10/04/P04015},
url = {https://dx.doi.org/10.1088/1748-0221/10/04/P04015},
year = {2015},
month = {apr},
publisher = {},
volume = {10},
number = {04},
pages = {P04015},
author = {P.D. Dauncey and M. Kenzie and N. Wardle and G.J. Davies},
title = {Handling uncertainties in background shapes: the discrete profiling method},
journal = {Journal of Instrumentation},
}

@article{diphoton,
    author = "{ATLAS Collaboration}",
    title="{Search for boosted diphoton resonances in the 10 to 70 GeV mass range using 138 fb$^{-1}$ of 13 TeV pp collisions with the ATLAS detector}",
   volume={2023},
   ISSN={1029-8479},
   url={http://dx.doi.org/10.1007/JHEP07(2023)155},
   DOI={10.1007/jhep07(2023)155},
   number={7},
   journal={Journal of High Energy Physics},
   publisher={Springer Science and Business Media LLC},
   year={2023},
   month=jul }

@misc{Frate2017ModelingSB,
  title="{Modeling Smooth Backgrounds and Generic Localized Signals with Gaussian Processes}",
  author={Frate, M. and Cranmer, K. and Kalia, S. and Vandenberg-Rodes, A. and Whiteson, D.},
  year={2017},
  eprint={1709.05681},
archivePrefix={arXiv},
primaryClass={physics.data-an},
howpublished="https://arxiv.org/abs/1709.05681", 
}

@article{scikit-learn,
    author = "{F. Pedregosa, G. Varoquaux, A. Gramfort, V. Michel, B. Thirion, O. Grisel, \textit{et. al.} }",
    year = {2011},
    publisher = {JMLR.org},
    volume = {12},
    number = {null},
    issn = {1532-4435},
    journal = {Journal of Machine Learning Research},
    pages = {2825–2830},
    title = {Scikit-learn: Machine Learning in Python},
}

@book{atlas_tdr,
    author = "{ATLAS Collaboration}",  
    collaboration = "ATLAS",
      title         = "{ATLAS: technical proposal for a general-purpose pp
                       experiment at the Large Hadron Collider at CERN}",
      publisher     = "CERN",
      address       = "Geneva",
      series        = "LHC technical proposal",
      year          = "1994",
      url           = "https://cds.cern.ch/record/290968",
      doi           = "10.17181/CERN.NR4P.BG9K",
}

@article{cms_tdr,
doi = {10.1088/0954-3899/34/6/S01},
url = {https://dx.doi.org/10.1088/0954-3899/34/6/S01},
year = {2007},
month = {apr},
publisher = {},
volume = {34},
number = {6},
pages = {995},
author = "{CMS Collaboration}",
title = "{CMS Physics Technical Design Report, Volume II: Physics Performance}",
journal = "{Journal of Physics G: Nuclear and Particle Physics}",
}

@article{ATLAS_Higgs,
title = "{Observation of a new particle in the search for the Standard Model Higgs boson with the ATLAS detector at the LHC}",
journal = "{Physics Letters B}",
volume = {716},
number = {1},
pages = {1-29},
year = {2012},
issn = {0370-2693},
doi = {https://doi.org/10.1016/j.physletb.2012.08.020},
url = {https://www.sciencedirect.com/science/article/pii/S037026931200857X},
author = "{ATLAS Collaboration}",
}

@article{CMS_Higgs,
   title="{Observation of a new boson at a mass of 125 GeV with the CMS experiment at the LHC}",
   volume={716},
   ISSN={0370-2693},
   url={http://dx.doi.org/10.1016/j.physletb.2012.08.021},
   DOI={10.1016/j.physletb.2012.08.021},
   number={1},
   journal="{Physics Letters B}",
   publisher={Elsevier BV},
   author="{CMS Collaboration}",
   year={2012},
   month=sep, pages={30–61} }

@article{Evans:2008zzb,
doi = {10.1088/1748-0221/3/08/S08001},
url = {https://dx.doi.org/10.1088/1748-0221/3/08/S08001},
year = {2008},
month = {aug},
publisher = {},
volume = {3},
number = {08},
pages = {S08001},
author = {Lyndon Evans and Philip Bryant},
title = "{LHC Machine}",
journal = {Journal of Instrumentation},
}

@book{GP,
    author =  "Rasmussen, C. and  Williams, C.",
    title = "{Gaussian Processes for Machine Learning}",
    publisher = "{The MIT Press}",
    year ="2005" 
}

@book{poissonProc,
    author = "J. F. C. Kingman",
    title = "{Poisson Processes}",
    publisher = "Clarendon Press",
    year = "1993"
}

@misc{barr2025gaussianprocessregressionsustainable,
      title="{Gaussian Process Regression as a Sustainable Data-driven Background Estimate Method at the (HL)-LHC}", 
      author={Barr, J. and Liu, B.},
      year={2025},
      eprint={2503.07289},
      howpublished = "{https://arxiv.org/abs/2503.07289}"   
}

@misc{gpytorch_lgcp,
  title = {Cox Processes (w/ Pyro/GPyTorch Low-Level Interface)},
  howpublished = {\url{https://docs.gpytorch.ai/en/latest/examples/07_Pyro_Integration/Cox_Process_Example.html}},
  note = {Accessed: 2025-07-17},
  year = 2023,
}

@article{Hayrapetyan_2024,
   title={Searches for Pair-Produced Multijet Resonances Using Data Scouting in Proton-Proton Collisions at $\sqrt{13}$~TeV},
   volume={133},
   ISSN={1079-7114},
   url={http://dx.doi.org/10.1103/PhysRevLett.133.201803},
   DOI={10.1103/physrevlett.133.201803},
   number={20},
   journal={Physical Review Letters},
   publisher={American Physical Society (APS)},
   author="{CMS Collaboration}",
   year={2024},
   month=nov }

@article{Gandrakota_2023,
   title={Model selection and signal extraction using Gaussian Process regression},
   volume={2023},
   ISSN={1029-8479},
   url={http://dx.doi.org/10.1007/JHEP02(2023)230},
   DOI={10.1007/jhep02(2023)230},
   number={2},
   journal={Journal of High Energy Physics},
   publisher={Springer Science and Business Media LLC},
   author={Gandrakota, Abhijith and Lath, Amit and Morozov, Alexandre V. and Murthy, Sindhu},
   year={2023},
   month=feb }

@article{Mathad_2021,
doi = {10.1088/1748-0221/16/06/P06016},
url = {https://dx.doi.org/10.1088/1748-0221/16/06/P06016},
year = "{2021}",
month = {jun},
publisher = {IOP Publishing},
volume = {16},
number = {06},
pages = {P06016},
author = {Mathad, A. and O'Hanlon, D. and Poluektov, A. and Rabadan, R.},
title = {Efficient description of experimental effects in amplitude analyses},
journal = {Journal of Instrumentation}
}

@article{Yallup2022HuntingFB,
    doi = {10.1088/1748-0221/18/05/P05014},
    url = {https://dx.doi.org/10.1088/1748-0221/18/05/P05014},
    year = {2023},
    month = {may},
    publisher = {IOP Publishing},
    volume = {18},
    number = {05},
    pages = {P05014},
    author = {Yallup, David and Handley, Will},
    title = {Hunting for bumps in the margins},
    journal = {Journal of Instrumentation},
}

@article{Bozson2018UnfoldingWG,
  title={Unfolding with Gaussian Processes},
  author={Adam James Bozson and Glen Cowan and Francesco Span{\`o}},
  journal={arXiv: Data Analysis, Statistics and Probability},
  year={2018},
  url={https://api.semanticscholar.org/CorpusID:53596609}
}

@article{Rodd2024CTAAS,
  title="{CTA and SWGO can discover Higgsino dark matter annihilation}",
  author={Nicholas L. Rodd and Benjamin R. Safdi and Weishuang Linda Xu},
  journal={Physical Review D},
  year={2024},
  url={https://api.semanticscholar.org/CorpusID:269983722}
}

@article{Foster2022ExtraterrestrialAS,
  title={Extraterrestrial Axion Search with the Breakthrough Listen Galactic Center Survey.},
  author={Joshua W. Foster and Samuel J. Witte and Matthew Lawson and Tim Linden and Vishal Gajjar and Christoph Weniger and Benjamin R. Safdi},
  journal={Physical review letters},
  year={2022},
  volume={129 25},
  pages={
          251102
        },
  url={https://api.semanticscholar.org/CorpusID:246904668}
}

@article{Abratenko2024FirstSM,
  title={First Simultaneous Measurement of Differential Muon-Neutrino Charged-Current Cross Sections on Argon for Final States with and without Protons Using MicroBooNE Data.},
  author={P. Abratenko and O. Alterkait and D. Andrade Aldana and L. Arellano and Jonathan Asaadi and Avi Ashkenazi and others},
  journal={Physical review letters},
  year={2024},
  volume={133 4},
  pages={
          041801
        },
  url={https://api.semanticscholar.org/CorpusID:268063188}
}

@misc{MicroBooNE2024MeasurementOT,
      title={Measurement of three-dimensional inclusive muon-neutrino charged-current cross sections on argon with the MicroBooNE detector}, 
      author="{MicroBooNE Collaboration}",
      year={2024},
      eprint={2307.06413},
      archivePrefix={arXiv},
      primaryClass={hep-ex},
      url={https://arxiv.org/abs/2307.06413}, 
}

@article{Hanuka2019OnlineTA,
  title={Online tuning and light source control using a physics-informed Gaussian process Adi},
  author={Adi Hanuka and Joseph Duris and Jane Shtalenkova and Dylan Kennedy and Auralee Edelen and Daniel Ratner and Xiaobiao Huang},
  journal={ArXiv},
  year={2019},
  volume={abs/1911.01538},
  url={https://api.semanticscholar.org/CorpusID:207780439}
}

@article{Stakia2021AdvancedMA,
  title={Advanced Multi-Variate Analysis Methods for New Physics Searches at the Large Hadron Collider},
  author={Anna Stakia and Tommaso Dorigo and Giovanni Banelli and Daniela Bortoletto and Alessandro Casa and Pablo de Castro and others},
  journal={ArXiv},
  year={2021},
  volume={abs/2105.07530},
  url={https://api.semanticscholar.org/CorpusID:234742087}
}

@article{Bertone2016AcceleratingTB,
  title="{Accelerating the BSM interpretation of LHC data with machine learning}",
  author={Gianfranco Bertone and Marc Peter Deisenroth and Jong Soo Kim and Sebastian Liem and R. Ruiz de Austri and Max Welling},
  journal={Physics of the Dark Universe},
  year={2016},
  url={https://api.semanticscholar.org/CorpusID:10754106},
  doi = {https://doi.org/10.1016/j.dark.2019.100293},
  volume = {24},
  pages = {100293},
}

@misc{Abratenko2023MeasurementOT,
      title={Measurement of three-dimensional inclusive muon-neutrino charged-current cross sections on argon with the MicroBooNE detector}, 
      author="{MicroBooNE Collaboration}",
      year={2024},
      eprint={2307.06413},
      archivePrefix={arXiv},
      primaryClass={hep-ex},
      url={https://arxiv.org/abs/2307.06413}, 
}

@article{Bertone_2018,
doi = {10.1088/1475-7516/2018/03/026},
url = {https://dx.doi.org/10.1088/1475-7516/2018/03/026},
year = {2018},
month = {mar},
publisher = {},
volume = {2018},
number = {03},
pages = {026},
author = {Bertone, Gianfranco and Bozorgnia, Nassim and Kim, Jong Soo and Liem, Sebastian and McCabe, Christopher and Otten, Sydney and de Austri, Roberto Ruiz},
title = {Identifying WIMP dark matter from particle and astroparticle data},
journal = {Journal of Cosmology and Astroparticle Physics}
}

@misc{Pappas2025HighFrequencyGW,
      title={High-Frequency Gravitational Wave Search with ABRACADABRA-10~cm}, 
      author="{K. M. W. Pappas, J. T. Fry, S. Cheng, A. C. Cesaní, J. L. Ouellet, C. P. Salemi, \textit{et. al.}}",
      year={2025},
      eprint={2505.02821},
      archivePrefix={arXiv},
      primaryClass={hep-ex},
      url={https://arxiv.org/abs/2505.02821}, 
}

@article{Dalmasso2020ConfidenceSA,
  title={Confidence Sets and Hypothesis Testing in a Likelihood-Free Inference Setting},
  author={Niccol{\`o} Dalmasso and Rafael Izbicki and Ann B. Lee},
  journal={ArXiv},
  year={2020},
  volume={abs/2002.10399},
  url={https://api.semanticscholar.org/CorpusID:211259400}
}

@article{Li2020EfficientNO,
  title={Efficient neutrino oscillation parameter inference using Gaussian processes},
  author={L. Li and Nitish Nayak and Jianming Bian and Pierre Baldi},
  journal={Physical Review D},
  year={2020},
  volume={101},
  pages={012001},
  url={https://api.semanticscholar.org/CorpusID:198938960}
}

@article{Kasieczka_2021,
doi = {10.1088/1361-6633/ac36b9},
url = {https://dx.doi.org/10.1088/1361-6633/ac36b9},
year = {2021},
month = {dec},
publisher = {IOP Publishing},
volume = {84},
number = {12},
pages = {124201},
author = {Kasieczka, Gregor and Nachman, Benjamin and Shih, David and Amram and others},
title = "{The LHC Olympics 2020 a community challenge for anomaly detection in high energy physics}",
journal = {Reports on Progress in Physics}
}

@article{Ablikim2023SearchFA,
  title={Search for a massless particle beyond the Standard Model in the $\Sigma^{+}\rightarrow p+$invisible decay},
  author="{BESIII collaboration}",
  journal={Physics Letters B},
  year={2023},
  url={https://api.semanticscholar.org/CorpusID:266573327},
  volume = {852},
  pages={138614},
}

\end{document}